\documentclass[]{tLCT2e}
\usepackage{graphicx}
\usepackage{subfigure}

\begin{document}

\title{Structure formation in monolayers composed of hard bent-core molecules}
\author{
Pawe\l{} Karbowniczek$^{\rm a}$, Micha\l{} Cie\'sla$^{\rm b}$ $^{\ast}$\thanks{$^\ast$Corresponding author. Email: michal.ciesla@uj.edu.pl \vspace{6pt}}, Lech Longa$^{\rm b}$ and Agnieszka Chrzanowska$^{\rm a}$ 
\\\vspace{6pt}
$^{a}${\em{Institute of Physics, Cracow University of Technology, ul.\ Podchor\c{a}\.{z}ych 1, 30-084, Krak\'{o}w, Poland.}}; $^{b}${\em{Marian Smoluchowski Institute of Physics, Department of Statistical Physics and  Mark Kac Center for Complex Systems Research,  Jagiellonian University, ul. \L{}ojasiewicza 11, 30-348 Krak\'{o}w, Poland.}}
}

\maketitle

\begin{abstract}
Two-dimensional ensembles of bent-core shaped molecules attain at highly orienting surfaces liquid crystalline structures characteristic mostly for lamellar chiral or nonchiral antiferroelectric order. Here, using the Onsager-type of density functional theory supplemented by constant-pressure Monte-Carlo (MC) simulation we investigate the role of excluded-volume interactions in stabilizing different structures in monolayers filled with bent-shaped molecules. We study influence of molecular features, like the apex angle, thickness of the arm and the type of the arm edges on the stability of layered structures. For simple molecular shapes taken the observed phases are dominated by the lamellar antiferroelectric type as observed experimentally, but a considerable sensitivity of the ordering to details of the molecular shape is found for order parameters and wave vectors of the structures. Interestingly, for large opening angles and not too thick molecules a window of stable nematic splay-bend  phase is shown to exist. The presented theory models equilibrium properties of bent-core liquid crystals subjected to strong planar anchoring,  in the case when details of the surface are of secondary importance.
\end{abstract}

%
\section{Introduction}
%
Two-dimensional structures made by complicated macromolecules are recently of great interest due to their potential applications, mainly in photoelectronic and biosensor area \cite{Gong,Matharu,Niu,Li,Iglesias,Banno,LiWalda,Querner}. In contrast to assemblies of spherical objects like, for instance, colloidal or nanosilica spheres, in case of anisotropic or irregularly shaped particles, there is a possibility to realize monolayers exhibiting very regular patterns which, next, can be utilized as a matrix capable to orient liquid crystal or to fabricate elements of electronic devices \cite{Iglesias}. It has also turned out that the structure of a matrix built within a monolayer may influence the activity of biomolecules. This biomolecular effect is a first step for biosensors construction. A comprehensive and detailed report about ordered molecular assemblies formed by Langmuir-Blodgett films and self-assemblies with potential influence on biosensing capabilities is given in \cite{Matharu}.

Achiral bent-core (banana) shaped molecules can be important in this regard \cite{Jakli,Takezoe,Gong,Niu,Li,Iglesias}. This arises from the observation of extraordinary self-organization in these mesogens in 3D, like the twist-bent nematic phase of nanoscale pitch \cite{Borshch,Chen}, the fibre forming smectic twist-bent phase  \cite{Tamba} and the cybotactic nematic phase \cite{Samulski}. They are also promising candidates to form the elusive biaxial nematic phase \cite{Grzybow}, the splay-bend nematic phase ($N_{SB}$) \cite{NSBa,NSBb,ShamidSB,LLSB} and even more complex structures with tetrahedratic order \cite{Lubensky,Pajak2}.

In 2D the situation is more subtle. These systems  are generally characterized by the lack of true long range order in the nematic state,  which is a consequence of director's fluctuations. A continuous nematic-isotropic phase transition  goes here via Kosterlitz-Thouless disclination unbinding mechanism yielding what is observed as algebraically decaying orientational pair correlation function in the nematic phase \cite{Kunz}. It is observed, for example, in simulations of a 2D system of hard needles with zero thickness \cite{FrenkelNeedle,Vink}, for planar hard rods \cite{Bates} and for zig-zag and bow-shaped hard needles \cite{2015-Stark}. Even though the true long-range nematic order does not exist in these systems on a macroscopic scale the simulations show that it persists over large spatial dimensions (\emph{i.e.} on a mesoscopic scale). Interestingly, it can be well described by means of the Onsager's Density Functional Theory (DFT) \cite{2000-Varga-Parsons-2dim,ActaAgCh}, despite the fact that macroscopic fluctuations of the director are generally not included in DFT.

On the experimental side the data of Gong and Wan \cite{Gong}
for banana-shaped liquid crystal molecules (P-n-PIMB) deposited
on a highly orienting pyrolytic graphite (HOPG) surface
reveal that the nematic order can be nearly saturated over the sample.
Using scanning tunneling microscopy (STM)
the authors observed here several antiferroelectric chiral and nonchiral lamellar structures.
Antiferroelectric smectic order in dense 2D banana systems
has been also discussed theoretically as prevailing
in \cite{BisiAFsmectic} by Bisi {\em et al.}  based on the packing arguments and, later, by Gonzales {\em et al.}  in the case of needle-like, infinitely thin boomerangs \cite{Gonzales} and hockey stick-shaped molecules consisting of two line segments  \cite{Gonzales2}. It has been also detected in
zero-thickness  zig-zag and bow-shaped systems \cite{2015-Stark}.
In addition,  in  \cite{Gonzales,Gonzales2,2015-Stark} the authors have observed that upon increasing pressure, before the system attains antiferroelectric smectic A phase,
a  spatially non-uniform, bend-deformed polar domains  
with the overall zero net polarization are being formed.

Understanding molecular self-organization in thin layers of more realistic, \emph{finite-thickness} bent-core molecules
is an interesting theoretical issue. Since most studies on two-dimensional systems are based on the particles exhibiting geometrical shapes like needles \cite{FrenkelNeedle,ActaAgCh},
hard discorectangles \cite{Martinez2005}, or zigzag particles \cite{Varga} interacting via hard core potentials, we will also incorporate a model from this class (it will be discussed in detail later). Both types of approach -  MC simulations
and Onsager's Density Functional Theory (DFT) - give  consistent predictions here. Of particular importance on the phase stabilization are excluded volume effects due to primary molecular features of the particles. In the case of bananas
these features are: length and width of the arms and the apex angle. As it will be shown, the secondary features
like \emph{e.g.} the shape of the arm's end contribute to 
quantitative changes.

As already mentioned above the DFT of Onsager's type has proven to give a good insight into qualitative features of the phases. One of the benefits of using the DFT scheme in connection
with bifurcation and symmetry analyses is the fact
that it allows to cover a broad range of cases giving clear directions for a more detailed study. The theory predicts the existence of the ordered mesophases with weakly first or second-order phase transitions in 2D systems, and hence cannot
predict quasi long-range order (QLRO), which is characteristic
for systems with a continuous broken symmetry. Even though the
Onsager's DFT does not account for QLRO, it works surprisingly well
for nonseparable, hard body interactions \cite{2000-Varga-Parsons-2dim,ActaAgCh}. Indeed, a comparison of the Onsager DFT with MC simulations suggest that the former theory is able to account for relevant features of molecular self-organization
\cite{Gonzales,FrenkelNeedle,2000-Varga-Parsons-2dim,ActaAgCh,Varga,2015-Stark}.

The aim of the present paper is to investigate with the Onsager's DFT a possibility of structure formation in monolayers
built from hard bent-core molecules of zero- and finite thickness. In particular we show that a change in molecular shape can have a profound effect on the properties and even stability of the structures. We mainly limit ourselves to the case of
high orientational order, in agreement with experiment  \cite{Gong} and previous 2D studies \cite{Gonzales}, but supplement the DFT analysis with a full Monte-Carlo simulation 
to support validity of the approximation used.

The paper is organized as follows:
Section II presents the model and Section III introduces
the Onsager's DFT formalism together with the appropriately identified order parameters, needed for the structure description of aligned boomerangs. Section IV gives the results of the bifurcation analysis for $N_{SB}$ and lamellar structures.
Section V provides exemplary phase diagrams obtained from the full minimization for three different bent-core systems:
hard needle-like bananas, finite thickness bananas with flat horizontal edges and finite thickness bananas with squared edges.
Finally, in the last section a summary is given  together with the main conclusions.
%
%
\section{Model}
\begin{figure}
\begin{center}
\includegraphics[width=0.7\columnwidth]{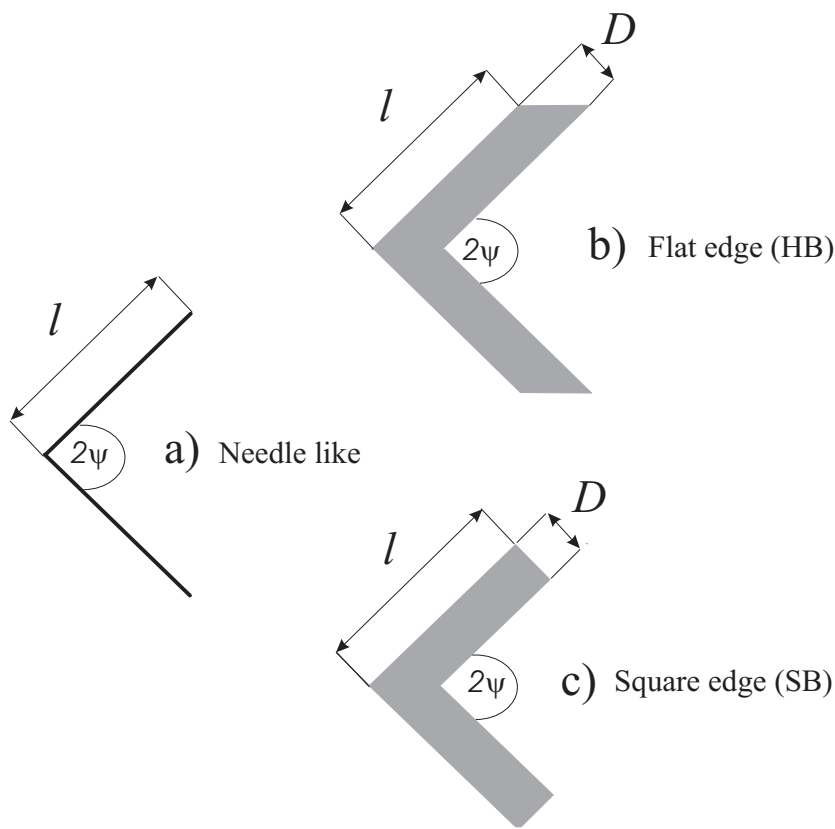}
\end{center}
\caption{Shapes of  bent-core molecules studied: (a) bent-core needles (they serve as a reference),
(b) finite thickness bent-core molecules  with flat horizontal edges (HB), and
(c) finite thickness bent-core molecules with squared
edges (SB). The apex angle, $\psi$, is here $\pi/4$ and the arm's width is $D=l/4$.}
\label{Fig01_trzyksztalty}
\end{figure}

We are going to study molecular self-organization in a two-dimensional system of hard bent-core
molecules of finite thickness.
Three types of molecules will be studied: needle-like boomerangs, Fig.~(\ref
{Fig01_trzyksztalty}a), finite thickness boomerangs with
horizontally cut edges (HB), Fig.~(\ref
{Fig01_trzyksztalty}b), and finite thickness boomerangs with squared edges (SB), Fig.~(\ref
{Fig01_trzyksztalty}c).

The bent core needles,  which are the reference particles given in Fig.~(\ref{Fig01_trzyksztalty}a), are just
two line segments of the length $l$ joined at one end in such a way that they form the apex angle of $2\psi$.
To obtain a HB molecule the line segments are replaced by rhomboids
whose tilt angle conforms to the assumption that the edges are effectively
horizontal as in Fig.~(\ref{Fig01_trzyksztalty}b).   The SB molecules  will differ from the
HB  molecules with respect to the shape of the arm edges, which in the SB case are squared.
$D$ describes here the arm's thickness.

We should add that we sought for several possibilities of introducing finite thickness to
needle-like boomerangs. The SB molecules seem to be the most natural extension,
whereas boomerangs with horizontally cut arms (the HB particles) are expected to attain
a layered arrangement more easily.  Indeed,  for the HB systems  even  close packing arrangements correspond to lamellar order with polarized layers.  Importantly, for all three cases the excluded areas can be calculated analytically. 

In order to compare the results for these three differently shaped bananas,
Fig.~(\ref{Fig01_trzyksztalty}), we introduce the dimensionless 
shape parameter  (width to arm's length ratio) $\delta=D/l$ ($0\le\delta\lesssim 1$)
 and define the reduced density in agreement with one used for bent-core needles \cite{Gonzales}
\begin{equation}
\rho=\bar{\rho}\, l ^2\sin(2\psi),
\label{rho_def}
\end{equation}
where ${\bar{\rho}}=N/S$ stands for the average density with
$N$ being the number of particles within the surface area $S$.
Using definition of the packing fraction parameter
$\eta = NS_{mol}/S$, with $S_{mol}$ being the surface of the molecule, 
the reduced density becomes
\begin{equation}
\rho=\eta \frac{l^2\sin(2\psi)}{S_{mol}}.
\label{rho_def1}
\end{equation}
In the case of the HB particles, $S_{mol}=2l^2\delta$. Thus
\begin{equation}
\rho= \eta \frac{\sin(2\psi)}{2 \delta}.
\label{rho_defB}
\end{equation}
For the SB particles $S_{mol}=l^2\delta\left(2-\delta/\tan(\psi)\right)$. Then
\begin{equation}
\rho=\eta \frac{\sin(2\psi)}{\delta(2-\delta/\tan (\psi)) } .
\label{rho_defC}
\end{equation}
Please note that the parametrization (\ref{rho_def1}) is singular for $\psi=0$ and $\psi=\pi/2$, where
bent-core molecules of zeroth thickness become reduced to a line, with $\eta=0$.
For 3D liquid crystals the typical packing fractions accessible to liquid crystalline
phases attain values from the interval (0.4-0.7).
For 2D systems, including $\delta=0$ case, $\eta$ spans
the whole interval (0-1). In particular,   HB and SB boomerangs 
can reach their maximal possible value of $\eta=1$
for ideal, close-packed, lamellar configurations
with maximally polarized layers. Very 
high packing fractions  for lamellar structures in 2D ($\eta \approx 0.8$)  
are also observed in experiments of Gong and Wan \cite{Gong}.

\section{Density functional analysis}

\subsection{Free energy functional and self-consistency equations}

A successful approach to describe the phase behaviour
of hard-body liquid crystalline systems is a generalization
of the Onsager theory. In this framework
the Helmholtz free energy, $\mathcal{F}$, is constructed
as a functional of the single particle
probability distribution function $P(X)$ \cite{DFTEvans}
\begin{eqnarray}
\frac{\mathcal{F}\left[ P \right]}{N k_B T}  &=&k_{B}T~\underset{(X)}{Tr}\left\{ \rho (X)%
\left[\, \ln \left( \Lambda \rho (X)\right) -1\right] \right\} +\underset{(X)}{%
Tr}\left[\, \rho (X)V_{ext}\right]   \nonumber  \\
&&-\frac{k_{B}T}{2}\underset{(X_{1},X_{2})}{Tr}\left[ \rho (X_{1})f_{12}\,\rho
(X_{2})\right],   \label{freeenergy}
\end{eqnarray}
where
\begin{equation}
f_{12}=e^{-\beta V(X_{1},X_{2})}-1  \label{Mayer}
\end{equation}
is the Mayer function.
Here $V$ is the interparticle potential,
$V_{ext}$ is the external potential, representing interaction
with external fields, or surfaces, $\Lambda$ is the constant resulting from the
integration over momenta, $T$
is the absolute temperature  and $k_B$ is the Boltzmann constant.
$\rho (X)$ stands for the
one-particle distribution function, which is normalized to the total number of
particles $N$
\begin{equation}
\underset{(X)}{Tr}[\rho (X)]=\left\langle N\right\rangle \equiv N.
\label{Trrho}
\end{equation}
In what follows no external
fields are taken into account and the surface is assumed smooth at the lengthscale
of the molecular size (typically a few nanometers for bent-core molecules).
Its role is limited to confine molecules in 2D (strong planar anchoring).
Under these assumptions the corresponding $V_{ext}$ does not depend on molecular
orientational degrees of freedom
and, hence, can be disregarded in the expansion (\ref{freeenergy}).

\begin{figure}[htb]
\begin{center}
\includegraphics[width=0.6\columnwidth]{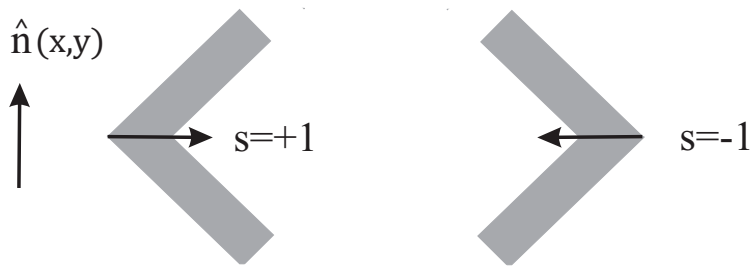}
\end{center}
\caption{{\protect\small {Definition of the $s$ variable, accounting for
different orientations of molecule's steric dipole with respect to the local director, $\hat{\mathbf{n}}(x,y)$. } }}
\label{Fig02_esy}
\end{figure}

The variable $X$ represents the set describing the position $ \mathbf{r}=(x,y)$
of the centre of mass of the particle and its orientations. In the description of
lamellar structures we assume, in agreement with experiment \cite{Gong}
and previous 2D studies \cite{Gonzales}, that orientational order is nearly saturated. In practice it means
that for a $C_{2h}$-symmetric molecule the orientational degrees of freedom become reduced to a discrete
variable, say $s$, accounting only for two possible orientations of the steric dipole ($s=\pm 1$)
with respect to the \emph{local} director ${\hat{\bf{n}}}({\bf r})$, where $s=+1$  denotes
a particle with a steric dipole pointing to the 'right' and $s=-1$ denotes a particle
with a dipole pointing to the 'left'. This means that the steric dipole is assumed to stay 
perpendicular to the local director, Fig.~(\ref{Fig02_esy}). 
In what follows we limit orientational degrees of freedom
of a molecule to the above two values, but carry out exemplary NPT MC simulation 
with full spectrum of translational and orientational degrees of freedom to check 
credibility of this approximation. 
Hence, in 2D the trace in Eqs.~(5,7) is calculated  as
\begin{eqnarray}
\underset{(X)}{Tr}=&&\sum_{s=\pm 1}\int_{0}^{L}dx\int_{0}^{L}dy= \nonumber \\
&& S\sum_{s=\pm 1}\frac{1}{L}\int_{0}^{Md=L}dy,
\end{eqnarray}
where $L$ represents the linear dimension of our sample ($S=L^2$);
$M$ stands for the number of layers and $d$
is the layer thickness in the case of smectics.

\begin{figure}[htb]
\begin{center}
\includegraphics[width=0.6\columnwidth]{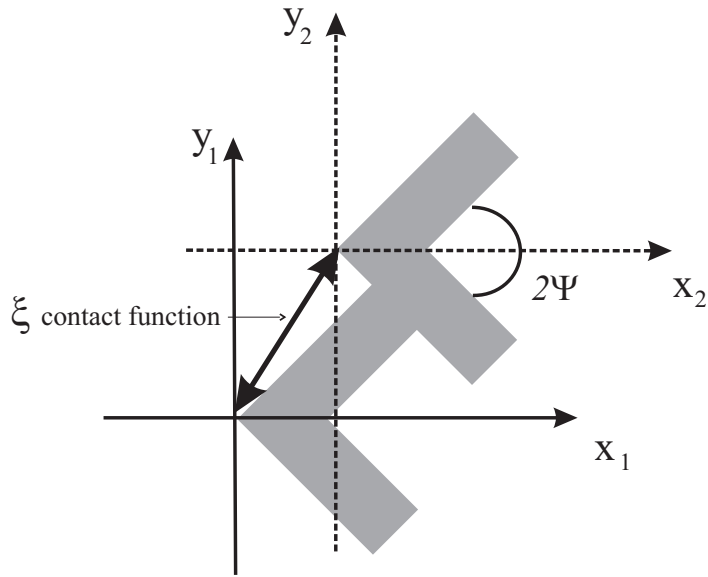}
\end{center}
\caption{{\protect\small {Definition of the contact function $\xi$
for two molecules with  the  apex angle $2\psi=\pi/2$ and $\delta=D/l=1/3$.  } }}
\label{Fig03_contact}
\end{figure}

In order to obtain the equilibrium solutions for the distribution function
the free energy functional $\mathcal{F}\left[ \rho \right] $ must be
minimized with respect to variation of $\rho(X)$ subject to the normalization
constraint $\underset{(X)}{Tr}[ \rho (X)]=N$.
It amounts in minimizing $\mathcal{F}^{\ast }\left[ \rho \right]$ given by
\begin{equation}
\mathcal{F}^{\ast }\left[ \rho \right] =\mathcal{F}\left[ \rho \right] -\mu
\left\{\underset{(X)}{Tr}[ \rho (X)]-N\right\},
 \label{Fstar}
\end{equation}
where $\mu$ is the chemical potential.
In our case of ideally oriented hard boomerangs the Mayer function
has a meaning of an excluded distance. It reads
\begin{equation}
f_{12}=e^{-\beta V(X_{1},X_{2})}-1=-\Theta \left[ \xi ({\hat{\mathbf
r}}_{12},s_1,s_2)-  {r}_{12}\right],  \label{f12teta}
\end{equation}
where ${\hat{\mathbf r}}_{12}=\frac{\mathbf{r}_{12}}{{r_{12}}}=
\frac{\mathbf{r}_{2}- \mathbf{r}_{1} }{|\mathbf{r}_{2}- \mathbf{r}_{1} |}$ and
$\xi$ is the contact function
defined as the
distance of contact from the origin of the second molecule for a given
direction ${\hat{\mathbf
r}}_{12}$ and orientations $s_1,s_2$
(see Fig.~(\ref{Fig03_contact}));
$\Theta$ denotes the Heaviside function.
Now, introducing the probability distribution function $P(X)$
\begin{equation}
\rho (X)=NP(X)=\overset{\_}{\rho }SP(X),
\end{equation}
and disregarding irrelevant (constant) terms, one can rewrite the free energy (5) in terms of
a rescaled free energy per unit area, $f(P)$, as
\begin{eqnarray}
&& \frac{f(P)}{l^2\sin(2\psi)}=\frac{\beta \Delta \mathcal{F}\left[ P\right] }{S}= \nonumber \\
&&
{\bar{\rho} }\,\underset{(X)}{Tr}\left[ P(X)\ln  P(X) \right] +\frac{%
{\bar{\rho }}}{2}\,\underset{(X)}{Tr}\left[ P(X)H_{eff}(X)\right],
\label{freeener}
\end{eqnarray}
where $H_{eff}$ is the effective excluded volume, averaged over probability distribution of particle "2". It reads
\begin{equation}
H_{eff}(X_{1})={\bar{\rho} }\,S\underset{(X_{2})}{Tr}\left\{
P(X_{2})\,\Theta \left[ \xi (X_{1},X_{2})-r_{12}\right] \right\}.
\label{Heff13}
\end{equation}
The equilibrium distribution function is now obtained by minimizing the free energy functional
(\ref{freeener}).
The necessary condition reads
\begin{equation}
\frac{\delta f(P)}{\delta P}=0,
\end{equation}
which in practice becomes reduced to solving  the self-consistent nonlinear integral equations
for $P(X)$
\begin{equation}
P(X)=Z^{-1}e^{-H_{eff}(X)},
\label{self1}
\end{equation}
where
\begin{equation}
Z=\underset{(X)}{Tr}e^{-H_{eff}(X)}
\label{self2}
\end{equation}
is the normalization of $P(X)$. The stationary solution of Eq.~(\ref{self1}) will be denoted $P_s(X)$.
\subsection{Details of the calculation}
\begin{figure}[htb]
\begin{center}
\includegraphics[width=0.75\columnwidth]{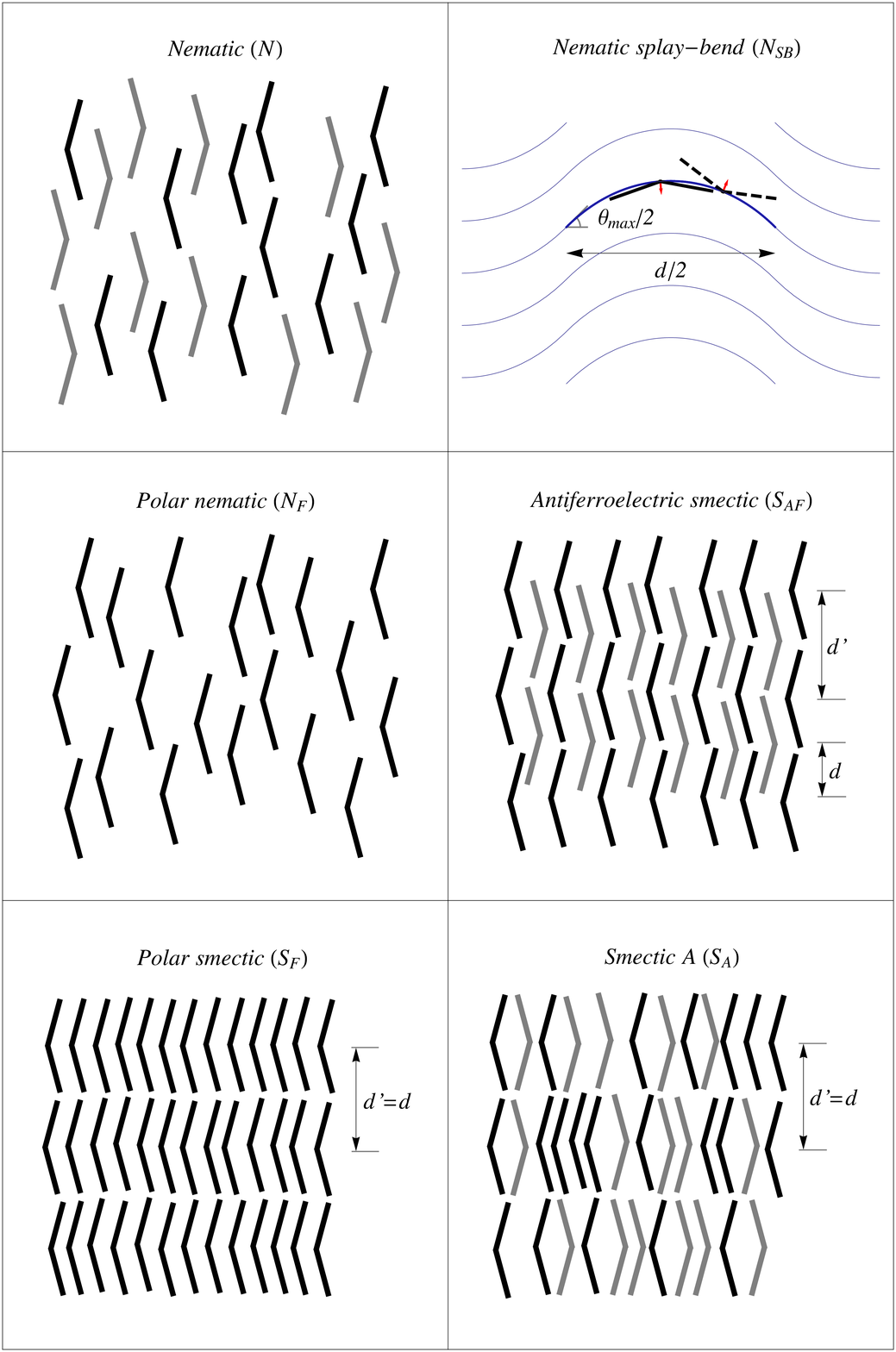}
\end{center}
\caption{{\protect\small {(Colour online) Possible arrangements of perfectly aligned
bent-core molecules. For better visibility the molecules  pointing in opposite direction are drawn
in different shades of gray. For the nematic splay-bend phase, which appears 
stable only for small $\delta$, molecules are represented by thick continuous- 
and dashed lines. The former corresponds to the most preferable orientation of the steric dipole (red arrow)
with respect to the local director while the later is less preferable orientation. Director is tangent 
to the lines shown.} }}
\label{Fig07_Allphases}
\end{figure}

In the analysis of possible stable phases we disregard phases with 2D periodicity,
like crystalline ones.  Structures that are left   
are polar nematics, $N_{SB}$  and  commensurate or incommensurate
smectics of A or C type, among which the most relevant  are  
shown in Fig.~(\ref{Fig07_Allphases}).
Thus, for the case of perfectly aligned boomerangs 
only two variables are needed to parametrize one particle distribution function. 
For  $N_{SB}$ we will assume the director to be a periodic function 
in x-direction, which means that $P(X)\equiv P(s,{\hat{\mathbf{n}}}(x))$,
where  $s=\pm1$ represents two opposite orientations 
of the steric dipole with respect to the local director, Fig.~(\ref{Fig07_Allphases}). 
For  lamellar structures we will use 
vertical coordinate $y$ and $s$ to parametrize $P$:
$P(X)\equiv P(s,y)$.
First, we will identify the bifurcation points
from the perfectly ordered reference nematic phase.

For the cases not involving $N_{SB}$ \cite{NSBa,NSBb}
the distribution function can be Fourier-expanded as
\begin{eqnarray}
 P(s,y)=
{\tilde{A}_{0}}+\sum_{n=1}^{\infty }{\tilde{A}_{n}}\cos
\left( \frac{2\pi ny}{d}-\phi_{0,n} \right) + &&\nonumber \\
s\,{\tilde{B}_{0}}+\sum_{m=1}^{\infty} s\,{\tilde{B}_{m}}\cos
\left( \frac{2\pi my}{d^{\,\prime }}-\phi_{1,m} \right),&& \;\; 0\le y \le L,
\label{Phi}
\end{eqnarray}
where periodic boundary conditions are assumed: $L=Md=M'd'$, with $M>0$ and $M'>0$ being integer numbers.

Note that the expansion (\ref{Phi}) is the most general representation for $P(s,y)$ 
 subject to periodic boundary conditions and given the structures are characterized  by 
 positionally independent, homogeneous  director field. It follows from the observation that $P(s,y)$,
 where $s = \pm 1$, is linear in $s$: $P(s,y)= \tilde{A}(y) + s \tilde{B}(y)$.
 Consequently, the independent Fourier expansions of $\tilde{A}(y)$ and
 $\tilde{B}$(y) involve the density wave part ($\tilde{A}_n$-terms) and
 the polarization wave part ($\tilde{B}_n$-terms) of periodicities $d$ and
  $d^{\,\prime}$, respectively. They are phase-shifted with respect to
  each other ($\phi$ phases), where the phases are determined up to
  a global phase, expressing freedom in choosing the origin of laboratory system of frame.

Using orthogonality properties of the Fourier series and properties of the s-space we can now define order parameters. They are given by
\begin{eqnarray}
&
\left\langle x_n \right\rangle =
\left\langle x \left(\frac{2\pi ny}{d}\right) \right \rangle &\nonumber \\
&
\left\langle sx_m \right\rangle =
\left\langle sx \left(\frac{2\pi my}{d'}\right) \right \rangle & ,
\label{csdef}
\end{eqnarray}
where
\begin{eqnarray}
\left\langle ... \right\rangle = \underset{(X)}{Tr} \left[ P(X)... \right ]
=S\sum_{s=\pm1}\frac{1}{L}\int_0^{Md=M'd'=L } dy P(s,y)...
\end{eqnarray}
with $x_{\alpha}\equiv\left\{ c,s \right \}$ and, correspondingly, $x(...)\equiv\left\{\cos(...),\sin(...) \right \}$.
With definitions (\ref{csdef}) we can finally rewrite the distribution function in the symmetry adapted form. It reads
\begin{eqnarray}
&&P(s,y)=\overset{}{\frac{1}{2S}}+\frac{1}{S}\sum_{n=1}^{\infty }
\left [
\langle c_{n} \rangle \cos \left( \frac{2\pi ny}{d} \right) + \langle s_n \rangle \sin \left( \frac{2\pi ny}{d} \right)
\right]  \nonumber \\
&&
+\frac{1}{2S} \langle s \rangle s
+\frac{1}{S}\sum_{m=1}^{\infty} \left[
\langle sc_{m} \rangle s\cos\left( \frac{2\pi ny}{d^{\prime }} \right)
+\langle ss_{m} \rangle s\sin \left( \frac{2\pi ny}{d^{\prime }} \right)
\right].
\label{Symadapted}
\end{eqnarray}
 Substituting the expansion (\ref{Symadapted}) back into the effective excluded volume
 (\ref{Heff13}) and assuming $L$ to be large, we can reduce $H_{eff}(X_1)$ to a simpler form. It is given by
\begin{equation}
H_{eff}(X_1)=\overset{-}{\rho} S \sum_{s_2=\pm 1} \int _0^{L} dy_2 \lambda (y_{12},s_{1}s_{2})
P(s_2,y_2),
\label{Heff21}
\end{equation}
where
\begin{equation}
\lambda (y_{12}, s_{1}  s_{2})
=\int_{0}^L \Theta \left[ \xi (x_{12},y_{12},s_{1} s_{2})-r_{12}\right] dx_{2}
=\lambda_0(y_{12})+s_1s_2\lambda_1(y_{12})
\label{lambda_df}
\end{equation}
plays the role of an excluded interval for fixed relative positions and orientations of two molecules.

\begin{figure}[htb]
\begin{center}
\includegraphics[width=0.5\columnwidth]{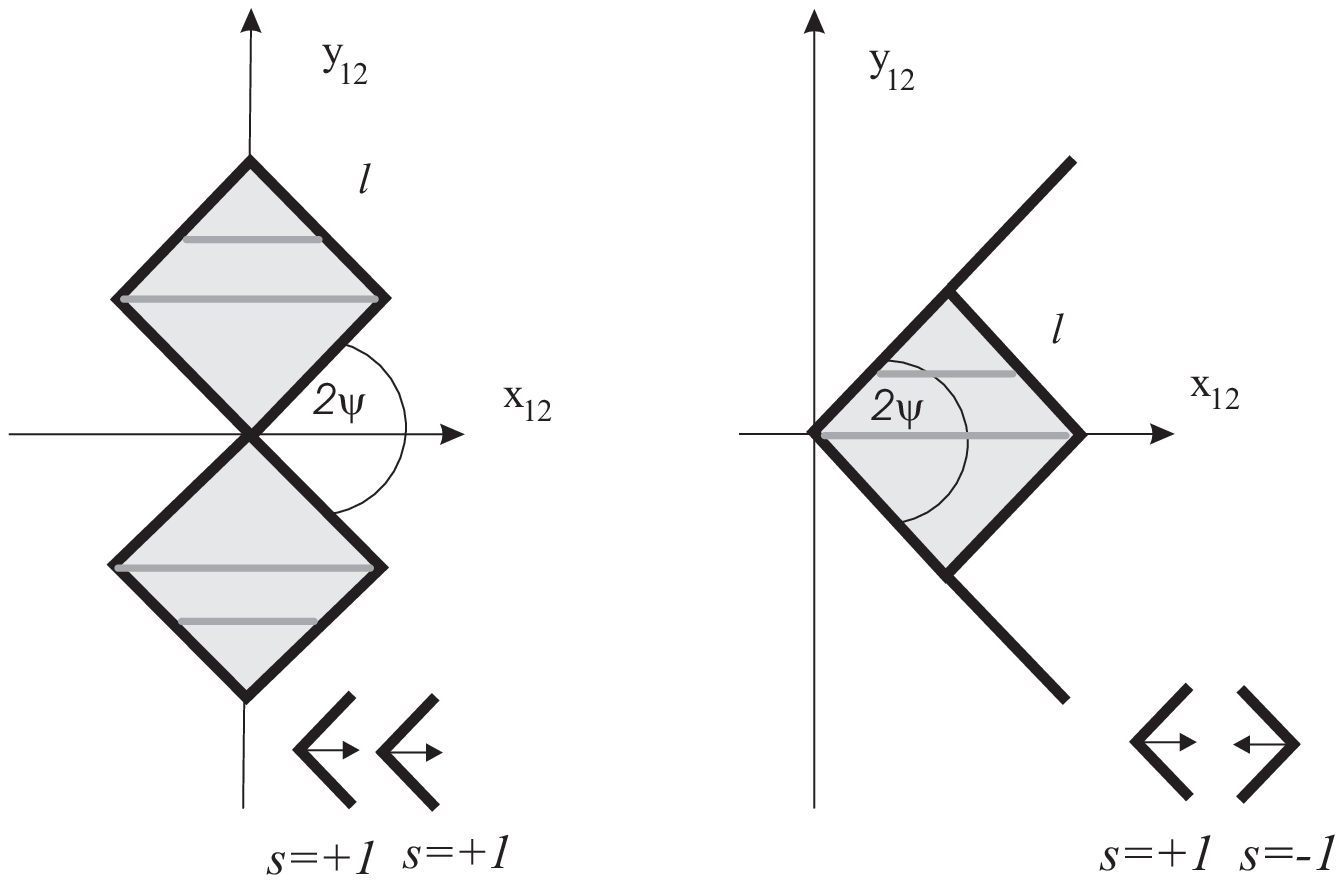} \\
\vspace{0.05\columnwidth}
\includegraphics[width=0.5\columnwidth]{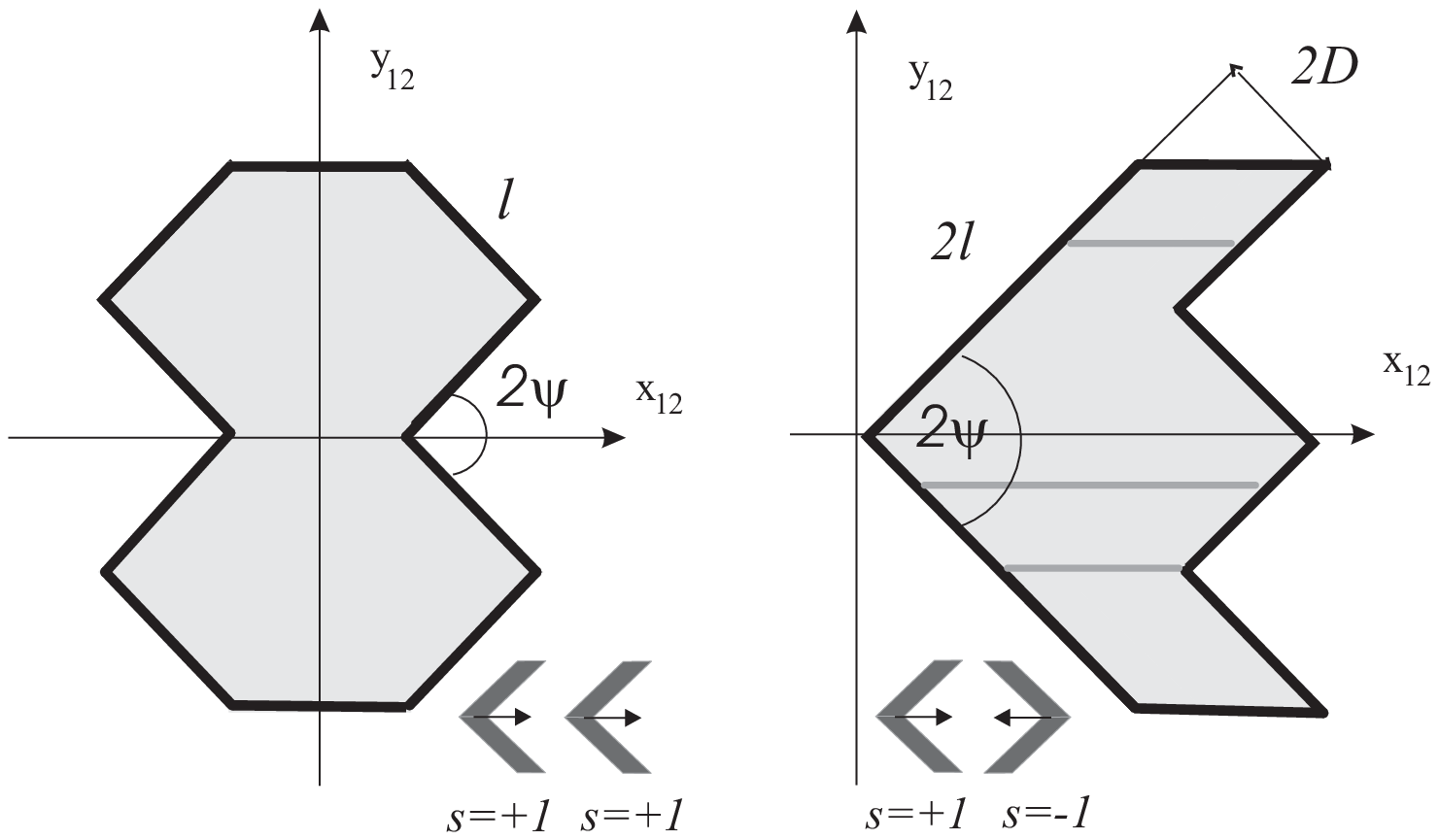}\\
\vspace{0.05\columnwidth}
\includegraphics[width=0.5\columnwidth]{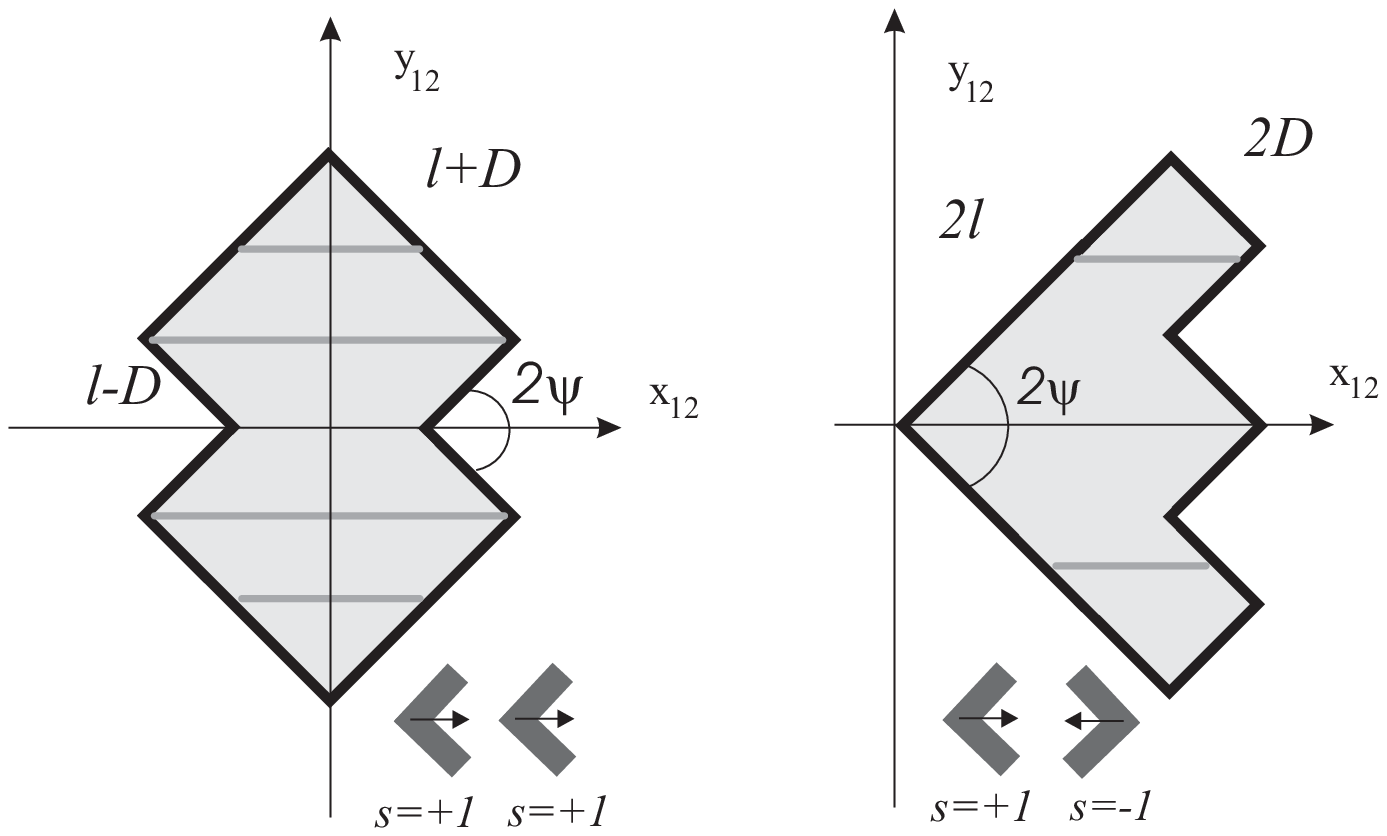}
\end{center}
\caption{{\protect\small {Excluded area in ($x_{12},y_{12}$) plane for needle-like, HB and SB particles with
$2\psi=\pi/2$ and $\delta=1/3$.  } }}
\label{Fig04_exclu}
\end{figure}

\begin{figure}[htb]
\begin{center}
\includegraphics[width=0.5\columnwidth]{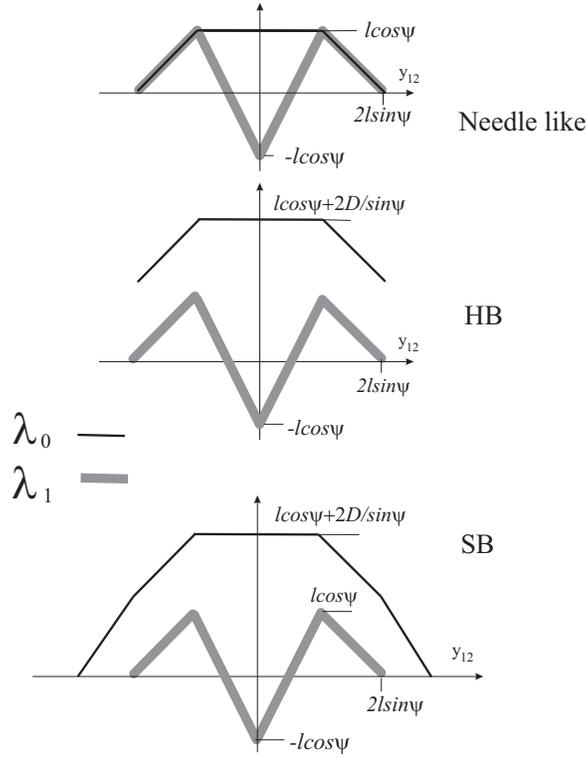}
\end{center}
\caption{{\protect\small { $\lambda$ functions for needle-like
 boomerangs, HB and SB molecules. Coordinates of characteristic points
of the functions are also given.} }}
\label{Fig05_lambdy}
\end{figure}

This excluded area depends only on the relative orientation between the molecules and on their relative distance.
 There are
two cases: with particles pointing in the same direction ($s_1s_2=1$), or in the opposite direction ($s_1s_2=-1$).
The exemplary cases are shown in Fig.~(\ref{Fig04_exclu}).
For our molecules the excluded area is calculated analytically, but only for the needle-like bananas
the formulas can be cast in a concise form. For $s_1s_2=1$ they read
%
\begin{eqnarray}
\left\{
\begin{array}{rllrllr}
-\frac{y_{12}+2l \sin \psi }{ \tan \psi }\leq & x_{12}& \leq \frac{y_{12}+2l \sin
\psi }{ \tan \psi } & \;\;\;\;\;\;-2l \sin \psi \leq &y_{12} &\leq -l\sin \psi \\
\frac{y_{12}}{\tan \psi }\leq & x_{12} &\leq -\frac{y_{12}}{\tan \psi } &\;\;\;\;\;\; -l\sin
\psi \leq &y_{12}& \leq 0 \\
-\frac{y_{12}}{ \tan \psi }\leq &x_{12} &\leq \frac{y_{12}}{\tan \psi }& \;\;\;\;\;\;\;\;\;\;\;\;0\leq&
y_{12} &\leq l\sin \psi \\
\frac{y_{12}-2l \sin \psi }{\tan \psi }\leq &x_{12}& \leq \frac{-y_{12}+2l \sin
\psi }{\tan \psi }& \;\;\;\;\;\;\;\;\;\;\;\;l \sin \psi \leq &y_{12} &\leq 2l \sin \psi
\end{array}\right.
\end{eqnarray}
and for $s_1s_2=-1$
\begin{eqnarray}
\left\{
\begin{array}{rllrllr}
&x_{12}&=-\frac{y_{12}}{\tan \psi } &\;\;\;\;\;\;-2l\sin \psi
\leq &y_{12}&\leq -l\sin \psi \\
-\frac{y_{12}}{\tan \psi }\leq &x_{12}& \leq \frac{y_{12}+2l\sin \psi }{\tan
\psi }  &\;\;\;\;\;\;-l\sin \psi \leq &y_{12} &\leq 0
\\
\frac{y_{12}}{\tan \psi }\leq &x_{12}& \leq \frac{-y_{12}+2l\sin \psi }{\tan
\psi } &\;\;\;\;\;\;\;\;\;\;\;\;0\leq &y_{12}&\leq l\sin \psi  \\
&x_{12}& \leq \frac{y_{12}}{\tan \psi } &\;\;\;\;\;\;\;\;\;\;\;\;l\sin \psi
\leq &y_{12} &\leq 2l \sin \psi.
\end{array} \right.
\end{eqnarray}
%
For needle-like boomerangs $\lambda$s take particularly simple form. They read
\begin{eqnarray}
&& \lambda _{0}(y_{12})=  \nonumber \\
&&
\left\{
\begin{array}{l}
l \cos \psi \\
\frac{-|y_{12}|+2l \sin \psi }{\tan \psi }
\end{array}
\right.
\begin{array}{r}
\;\;\;\;\;\;|y_{12}|<l \sin \psi \\
\;\;\;\;\;\;l \sin \psi \leq |y_{12}|\leq 2l \sin \psi,
\end{array}
\nonumber \\
\end{eqnarray}
\begin{eqnarray}
&&
\lambda _{1}(y_{12})= \nonumber\\
&&
\left\{
\begin{array}{l}
\frac{2|y_{12}|-l \sin \psi }{\tan \psi } \\
\frac{-|y_{12}|+2l \sin \psi }{\tan \psi }
\end{array}
\right.
\begin{array}{r}
\;\;\;\;\;\;|y_{12}|<l \sin \psi \\
\;\;\;\;\;\; l \sin \psi \leq |y_{12}|\leq 2l\sin \psi.
\end{array}
\nonumber \\
\end{eqnarray}
 Examples of $\lambda$s for needle-like boomerangs, HB and SB molecules are shown in Fig.~(\ref{Fig05_lambdy}).

The next step is to perform the integration in (\ref{Heff21}) by replacing $y_{12}$ with $y$, where $y_2=y_1+y$.
In the limit of large $L$ the final formula for the effective excluded volume is given by
\begin{eqnarray}
l^2 \sin(2\psi)H_{eff}(s_1,y_{1}) &=&{\rho }A_{0}+{\rho }B_{0} \langle s\rangle s_{1}
+
\nonumber \\ &&
2{\rho }\sum_{n=1}^{\infty }
\left\{ \langle c_{n} \rangle \left[ A_{n}\cos \left( \frac{2\pi
ny_{1}}{d} \right) - C_{n}\sin \left( \frac{2\pi
ny_{1}}{d} \right) \right] \right.+
\nonumber \\ &&
\left.
 \langle s_{n} \rangle \left[ A_{n} \sin \left( \frac{2\pi
ny_{1}}{d} \right) + C_{n} \cos \left( \frac{2\pi
ny_{1}}{d} \right) \right]
\right \}+
\nonumber \\ &&
2{\rho }s_{1}\sum_{m=1}^{\infty }
\left\{ \langle sc_{m} \rangle \left[ B_{m}\cos \left( \frac{2\pi
 my_{1}}{d^{\prime }} \right) -
D_{m}\sin
\left( \frac{2\pi my_{1}}{d^{\prime }} \right) \right] +\right.
\nonumber \\ &&
\left. \langle ss_{m} \rangle \left[ B_{m}\sin \left( \frac{2\pi
 my_{1}}{d^{\prime }} \right)+
D_{m}\cos
\left( \frac{2\pi my_{1}}{d^{\prime }} \right) \right] \right \},\;
\end{eqnarray}
where the coefficients of the expansion are defined as
\begin{eqnarray}
A_{n}=l^2 \sin(2\psi) \,\alpha(\psi,\delta,k_n)=
\int_{-2l \sin\psi}^{2l\sin\psi}\lambda_0(y)\cos\left( \frac{2\pi ny}{d}\right)dy,
\label{An}
\end{eqnarray}
\begin{eqnarray}
B_{m}=l^2 \sin(2\psi) \,\beta(\psi,\delta,k'_m)=
\int_{-2l \sin\psi}^{2l\sin\psi}\lambda_1(y)\cos\left( \frac{2\pi my}{d'}\right)dy,
\label{Bm}
\end{eqnarray}
\begin{eqnarray}
C_{n}=l^2 \sin(2\psi) \,\gamma(\psi,\delta,k_n)=
\int_{-2l \sin\psi}^{2l\sin\psi}\lambda_0(y)\sin \left( \frac{2\pi ny}{d}\right)dy,
\label{Cn}
\end{eqnarray}
\begin{eqnarray}
D_{m}=l^2 \sin(2\psi) \,\sigma(\psi,\delta,k'_m)=
\int_{-2l \sin\psi}^{2l\sin\psi}\lambda_1(y)\sin \left( \frac{2\pi my}{d'}\right)dy.
\label{Dm}
\end{eqnarray}
Here $\rho$ is defined in Eq.~(\ref{rho_def}) and $k$, $k'$ are dimensionless wave vectors given by
$k=\frac{\pi l \sin\psi}{d}$ and $k\,'=\frac{\pi l \sin\psi}{d'}$, respectively.
As a result of the condition that $L=Md=M\,'d\,'$ we additionally have a limitation $Mk\,'=M\,'k$ imposed on wave vectors.
Substitution of $n=m=0$ in (\ref{An}) and (\ref{Bm}) gives $A_0$ and $B_0$.

\begin{figure}[htb]
\begin{center}
\includegraphics[width=0.45\columnwidth]{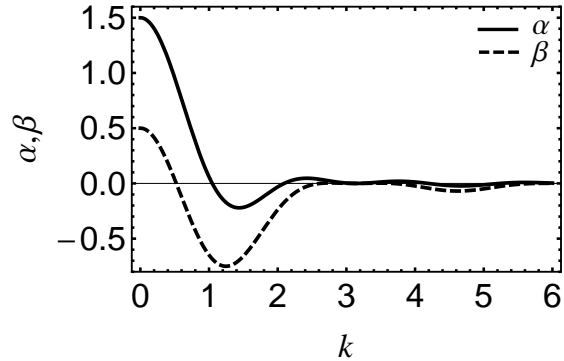}
\end{center}
\caption{{\protect\small {The k-dependence of the coefficients $\alpha$ and $\beta$ for needle-like boomerangs.}  }}
\label{Fig06_ABem}
\end{figure}

Formally, the coefficients (\ref{An}-\ref{Dm}) are the Fourier transforms of $\lambda_{\alpha}$, Eq.~(\ref{lambda_df}).
For the case of needle-like bananas these coefficients are of particularly simple form, namely
\begin{eqnarray}
\alpha(\psi,\delta,0)=\frac{3}{2},\; \beta(\psi,\delta,0)=\frac{1}{2},\;\gamma(\psi,\delta,k)=\sigma(\psi,\delta,k)=0, \nonumber\\
\alpha(\psi,\delta,k)=\frac{[2\cos(2k)+1]\sin^2(k)}{2k^2},\nonumber\\
\beta(\psi,\delta,k)=\frac{[2\cos(2k)-1]\sin^2(k)}{2k^2},
\label{alfy}
\end{eqnarray}
where  $k=\frac{l\pi\sin\psi}{d}$.
They are shown in Fig.~(\ref{Fig06_ABem}).

Taking definitions (\ref{csdef})  the nonlinear integral equation (\ref{self1})
becomes reduced to an infinite set of nonlinear algebraic equations for the order parameters
\begin{equation}
\left(
\begin{array}{c}
\langle x_{n} \rangle \\
\langle sx_{m} \rangle \\
\langle s \rangle
\end{array}
\right) =Z^{-1}\underset{(X_1)}{Tr} \left[ \left(
\begin{array}{c}
x \left( \frac{2\pi ny_{1}}{d} \right) \\
s_{1}x \left( \frac{2\pi m y_{1}}{d^{\prime }} \right) \\
s_{1}
\end{array}
\right) \exp (-H_{eff})\right].
\label{self34}
\end{equation}
The corresponding stationary excess free energy in the limit of large $L$ is given by
\begin{eqnarray}
\frac{f\left[ P_{s}\right]}{l^2 \sin(2\psi)} &=&{\bar{\rho} } \langle \ln P_{s} \rangle +\frac{{%
\bar{\rho} }}{2} \langle H_{eff} \rangle =-\frac{{\bar{\rho}}}{2} \langle H_{eff} \rangle
-{\bar{\rho} }\ln Z \nonumber \\
&&=-\frac{{\bar{\rho}}^2 }{2} \left[  A_{0}+B_{0} \langle s \rangle^{2}+2\sum_{n=1}^{%
\infty }A_{n} \left( \langle c_{n}\rangle^{2}+ \langle s_n\rangle^2 \right) \right. \nonumber \\
&& \left.
+2 s_1 \sum_{m=1}^{\infty }B_{m} \left( \langle sc_{m}\rangle^{2}+ \langle ss_m\rangle^2 \right) \right] -%
{\bar{\rho} }\ln Z.
\label{excess35}
\end{eqnarray}

Similar to the previous cases we can now proceed with the analysis of $N_{SB}$.
For the $N_{SB}$ structure  we take generalization of  one
proposed earlier by Meyer \cite{NSBa}. Since the steric polarization is always perpendicular
to the local director  we assume direction of the former to rotate uniformly between
 $-\theta_{max}/2 $ and $\theta_{max}/2 $  for $0\le x <d/2$, Fig.~(\ref{Fig07_Allphases}),
 where $\theta_{max}$ and $d$ should be determined from the free energy minimum.
Assuming the structure to be globally nonpolar we take
\begin{equation}\label{PNSB}
  P(s,{\mathbf{\hat n}}(x)) =  P(s,{\mathbf{\hat n}}(x+d)) = \frac{1}{2 S} +
  \frac{1}{2S} \langle s \rangle s \delta[{\mathbf{\hat{s}}}(x) +\{ \sin(\phi(x)), \cos(\phi(x))\} ],
\end{equation}
where
\begin{equation}
\phi(x) =
\left\{
\begin{array}{c}
\frac{2 \theta_{max} x}{d}- \frac{\theta_{max}}{2},  \hspace{2.7cm}  0 \le x < d/2   \\
\pi -\frac{ 2 \theta_{max} x}{d}+ \frac{3\theta_{max}}{2},  \hspace{1cm}  d/2 \le x < d.
\end{array}
\right.
\label{NSBphi}
\end{equation}
Here $s {\mathbf{\hat{s}}}(x)$ is the local polarization ($ {\mathbf{\hat{s}}}(x) \perp  {\mathbf{\hat{n}}}(x)$) with $s=\pm1$, as previously, and  $\delta[...]$ is the Dirac delta function. Note that the $N_{SB}$ structure needs  $d$, $\theta_{max}$, and the average polarization $<s>$ measured with respect to the local director to be determined variationally from the free energy  $f[P_s]$.  
The calculation of  the equations for $\langle s \rangle$ and for $f[P_s]$ goes similar as previously.  We only need to
interchange $y$ with $x$ in formulas (\ref{Heff21}, \ref{lambda_df}, \ref{self34}, \ref{excess35}), set $M=1$ and disregard
$M'$. The final formulas are identical to (\ref{self34}, \ref{excess35}) with $A_n=B_n=0$  for $n\ge1$.
The only difference now is that  $A_0$  and $B_0$ depend on $\theta_{max}$ and $d$.

The Eqs (\ref{self34}) should be solved for given $d,d'$ and appropriately chosen $M,M'$.
Then the equilibrium structure  is identified
with the absolute minimum of (\ref{excess35})
taken with respect to the stationary solutions and with respect
to the periodicities $d,d'$ and $\theta_{max}$. Note that the trivial nematic state corresponding to
$\langle s \rangle =\langle c_n \rangle =\langle s_n \rangle = \langle sc_n \rangle = \langle ss_n \rangle =0$
($\forall$ $n$) always satisfies Eqs (\ref{self34}).
Problem that remains is  to identify all nontrivial solutions of Eqs.~(\ref{self34}), where at least
one of the order parameters becomes nonzero. A systematic way of finding these solutions
is bifurcation analysis \cite{LongaRomano}.
Here, we apply this technique to analyse bifurcation from the nematic phase. We also determine
exemplary phase diagrams from the full minimization of
the free energy in different phases,
in a wide range $\eta$. We also carry out exemplary MC simulations that involve a full spectrum of orientational degrees of freedom  to support usefulness  of the ideal nematic order approximation. 
 \subsection{Bifurcation analysis}
Now we consider a bifurcation from a perfectly aligned nematic phase. Close to the bifurcation point
the difference between the states is arbitrarily small for each $d,d'$, which enables one to linearize
the RHS of Eqs.~(\ref{self34}) with respect to the order parameters. The analysis is carried out
by taking the needle-like boomerangs as reference. The results are
\begin{equation}
\left(
\begin{array}{c}
\langle c_{n} \rangle \\
\langle s_{n} \rangle \\
\end{array}
\right) =-\rho {\bf A}_n  \left(
\begin{array}{c}
\langle c_{n} \rangle \\
\langle s_{n} \rangle \\
\end{array}
\right),
\label{36a}
\end{equation}
\begin{equation}
\left(
\begin{array}{c}
\langle sc_{m} \rangle \\
\langle ss_{m} \rangle \\
\end{array}
\right) =-\rho {\bf B}_m  \left(
\begin{array}{c}
\langle sc_{m} \rangle \\
\langle ss_{m} \rangle \\
\end{array}
\right),
\label{36b}
\end{equation}
\begin{equation}
\langle s\rangle =\rho \beta(0) \langle   s \rangle,
\label{36c}
\end{equation}
where $2$ by  $2 $ arrays ${\bf A}_n$, ${\bf B}_m$  are given by
\begin{equation}
{\bf A}_n = \left(
\begin{array}{c}
\;\;\alpha(\psi, \delta, k_n) \;\; \gamma(\psi, \delta, k_n) \\
-\gamma(\psi, \delta,k_n) \;\; \alpha(\psi, \delta,k_n) \\
\end{array}
\right)
\label{37a}
\end{equation}
and
\begin{equation}
{\bf B}_m = \left(
\begin{array}{c}
\;\;\beta(\psi, \delta, k'_m) \;\; \sigma(\psi, \delta, k'_n) \\
-\sigma (\psi, \delta, k'_m) \;\; \beta(\psi, \delta, k'_n) \\
\end{array}
\right).
\label{37b}
\end{equation}
The homogeneous equations (\ref{36a}-\ref{36c}) have a nontrivial solution given that
at least one of the equations
\begin{equation}
\det({\bf 1}+\rho {\bf A}_n)=0,
\label{38a}
\end{equation}
\begin{equation}
\det({\bf 1}+\rho {\bf B}_m)=0,
\label{38b}
\end{equation}
\begin{equation}
\rho \equiv \rho_0=-\frac{1}{\beta(0)}
\label{38c}
\end{equation}
is satisfied for a positive $\rho$.
By solving (\ref{38a}, \ref{38b}) for $\rho$ we obtain two functions:
$\rho(k_n)$ and $\rho'(k'_m)$ , respectively, together with $\rho_0$.
The bifurcation density is then
identified with the lowest positive
value taken out of
\begin{equation}
\left \{ \underset{\{k_n\}}{\rm Min}\left[\,\rho(k_n)\,\right],
\underset{\{k'_m\}}{\rm Min}\left[\, \rho'(k'_m)\,\right],\rho_0
\right\}.
\label{39}
\end{equation}
For majority of cases studied we will assume the director to be perpendicular
both to the molecule's dipole moment and the layer normal, Figs.~(\ref{Fig02_esy}).
In this case $\lambda=\lambda(|y_{12}|,s_1s_2)$ in Eq.~(\ref{lambda_df}).
Consequently, we can choose $\phi s$ in (\ref{Phi}) to vanish and consider the case of vanishing
$\langle s_n \rangle$ and $\langle ss_m \rangle$. The corresponding bifurcation density is then the lowest
positive value out of
\begin{equation}
\left \{ \underset{\{k_n\}}{\rm Min}\left[\,\frac{-1}{\alpha(k_n)}\,\right],
\underset{\{k'_m\}}{\rm Min}\left[\, \frac{-1}{\beta(k\,'_m)}\,\right],
\frac{-1}{\beta(0)}
\right\}.
\label{40}
\end{equation}

As an example we start with the discussion of bifurcation for needle-like boomerangs.
It turns out that the most stable structures bifurcating from the nematic phase is
the antiferroelectric smectic A phase ($S_{AF}$). To see this consider the behaviour of
$\alpha(k_n)$  and $\beta(k'_m)$, shown in Fig.~(\ref{Fig06_ABem}).
One observes that $\beta$  attains the absolute minimum for $k_{0,m}^{\prime }=1.246$ and
$\alpha$ for $k_{0,n}=1.438$, and that these points correspond to the layer thicknesses of
$d_{0,m}^{\,\prime }=\frac{l m  \pi \sin \psi }{k_{0,m}^{\prime }}=2.52 m l \sin
\psi $ and $d_{0,n}=\frac{l n  \pi \sin \psi }{k_{0,n}^{{}}}=2.184 n l \sin
\psi $, respectively. Consequently, the physical bifurcation to the smectic A ($S_A$)
phase should occur for $n=1$ with $%
d_{0}\approx 2.18l \sin (\psi )= 1.09(2 l \sin (\psi ))$
and bifurcation to the smectic AF ($S_{AF}$)  phase for $m=1$ with $%
d_{0}\approx 2.51\sin (\psi )= 1.26(2l \sin (\psi ))$.
Since the minimum of $B_{m}$ is deeper, the expected lamellar phase bifurcating from the nematic phase will be of the
antiferroelectric type.
 The value of this minimum
determines the bifurcation density $(\rho_{bif}=-\frac{1}{B_{1}^{\min }})$.
Then,  the distribution function at the bifurcation point will take the form
\begin{equation}
P(s,y)= \frac{1}{2S}+\frac{1}{S}\langle sc_{1} \rangle s \cos \left( \frac{2\pi y}{%
2.52l \sin \psi } \right) +...
\end{equation}

\section{Possible structures}

In general, the most probable two-dimensional structures that can be expected in boomerang systems are given in Fig.~(\ref{Fig07_Allphases}). The low density phase, the nematic phase, can be here of two types: the standard nematic phase ($N$) in which the same (on average) number of boomerangs is pointing to the right as to the left. When one type of the orientation prevails ($<s> \neq 0$) then one deals with the polar (ferroelectric) nematic ($N_F$). One may also expect that the $N_{SB}$ phase ($<s> \neq 0$, $\text{finite}\,\, d$, $0<\theta_{max}\le \pi$) should be at least locally stable. Upon an increase of density a transition to a smectic phase, which is characterized by a regular modulation of the density profile due to presence of the layers, may occur. Three different smectic phases are plausible: the typical smectic A phase ($S_A$), where left and right-pointing particles are, on average, equally distributed ($<c_n> \ne 0$), the ferroelectric smectic A phase ($S_F$)  when the particles oriented in one direction overwhelm the number of the oppositely oriented particles ($<s>\ne 0$, $<c_n> \ne 0$, $<sc_n> \ne 0$, $d=d'$), and the antiferroelectric smectic phase ($S_{AF}$), in which the particles in subsequent layers have opposite orientations ($<s c_n> \ne 0$, $<c_n> \ne 0$, $d'=2d$). Note that the period $d\,'$ of the layers with particles  of  the same average orientation in the antiferroelectric phase is twice the smectic period $d$ ($M=2M'=2$), whereas in the polar phase they attain the same value. The occurrence of such phases will depend on the structure of the particles themselves as well as on the density.  

\section{Boomerangs of arms with finite width}
\subsection{The HB molecules}
The model of the needle-like boomerangs can be extended
to the case when arms are of finite width in many different
ways of which we choose HB and SB shapes.
The HB case is given in
Fig.(\ref{Fig01_trzyksztalty}b).
%
For HB molecules the coefficients $\alpha$ and $\beta$ are given by
\begin{eqnarray}
&& \alpha(\psi, \delta, k)=\frac{\left[1+2\cos\left( 2k \right )\right ]\sin^2k+\frac{4\delta}{\sin(2\psi)}k \sin\left(4k \right)}{2k^2},
\nonumber \\
&&  \beta(\psi, \delta, k)=\frac{\left[-1+2\cos\left( 2k \right )\right ]\sin^2k}{2k^2},
\nonumber \\
&& \alpha(\psi, \delta, k\rightarrow 0)=\frac{3}{2}+\frac{8 \delta}{\sin(2\psi)},
\nonumber \\
&& \beta (\psi, \delta, k\rightarrow 0)=\frac{1}{2}.
\end{eqnarray}
 \begin{figure}[htb]
\begin{center}
\vspace{0.8cm} \includegraphics[width=8.0cm,scale=1.0]{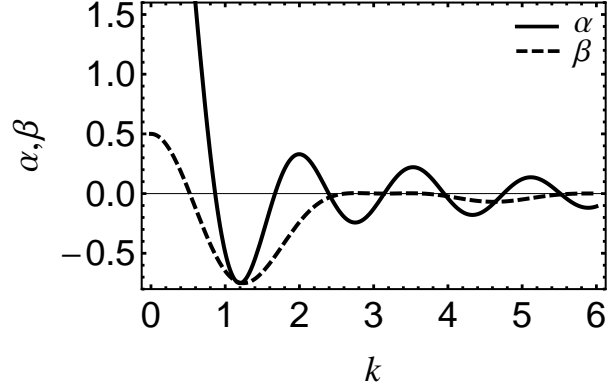}
\end{center}
\caption{{\protect\small {
The Fourier transforms $\alpha$ and $\beta$ for the condition $\delta=0.3654 \sin(2\psi). $}}}
\label{Fig08_alfabeta_fat}
\end{figure}
Note that the coefficient $\beta$ is here exactly the same as in the case of the needle-like boomerangs.
It turns out, however, that  when the condition
$\delta=0.3654\sin(2\psi)$ is fulfilled, the minimum of $\alpha$ and the minimum of $\beta$ attain the same value
$\beta_{min}=\alpha_{min}=-0.749956$  (\emph{see} Fig.~(\ref{Fig08_alfabeta_fat})).
This condition provides a set of values for the parameters serving as a limiting
case when the bifurcation from $N$ to  $S_A$ or  $S_F$  is observed. Note that the close packed structure  for HB molecules is of  lamellar  (smectic A) type with maximally polarized layers, but the direction of polarization within  
the layer is  doubly degenerate (${\mathbf{\hat{s}}} = \pm {\mathbf{\hat{x}}}$).  

\begin{figure}[htb]
\begin{center}
\includegraphics[width=0.6\columnwidth]{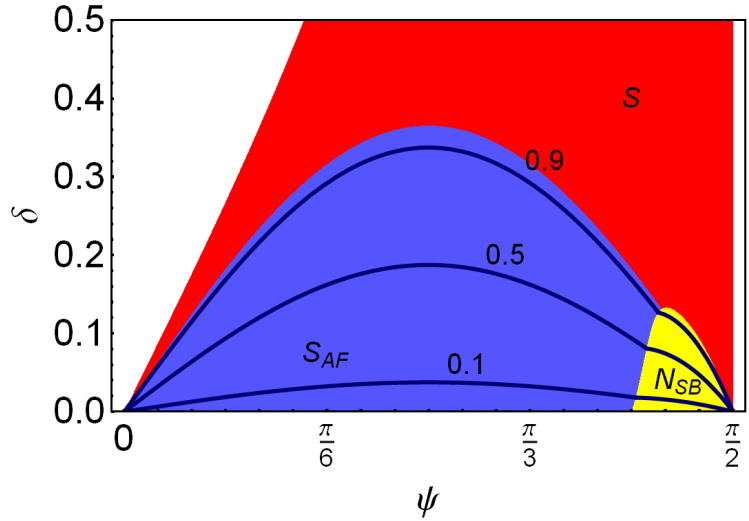}
\end{center}
\caption{{\protect\small {(Colour online) Bifurcation diagram for HB boomerangs. The blue region corresponds to the cases where
nematic ($N$)-antiferroelectric smectic A ($S_{AF}$)  bifurcation takes place and the red region  ($S$) corresponds to the cases where the
transition from $N$ undergoes either to $S_{A}$ or $S_{F}$. In the yellow region  the $N_{SB}$ structure
bifurcates from $N$. The black lines are the density limits with the  packing fraction $\eta$  given in the legend.
} }}
\label{Fig09_bifuPawel}
\end{figure}

Using  the density of the form $\rho=\eta \frac{ \sin 2\psi}{2\delta}$
one can now obtain the bifurcation diagram as given in Fig.~(\ref{Fig09_bifuPawel}).  The lines provide
condition where the normalized packing fraction $\eta$  is equal to $0.1$, $0.5$ and $0.9$.
Similar to \cite{Gonzales} the most common smectic phase  obtained here is $S_{AF}$,  given by the blue region. 
In the red region ($S$) of Fig.~(\ref{Fig09_bifuPawel}) the bifurcation scenario  leads to a single 
amplitude $S_A$ structure  ($<c_1> \ne 0$).  However, due to the 
coupling of $<c_1>$ with $<s>$ and $<sc_1>$ the equilibrium lamellar structures with polarized layers, like
$S_F$, are also  possible.    They can be stabilized  as a  result of  a phase transition between two different lamellar
structures and  cannot  be obtained by studying bifurcation from  $N$.  We would need to solve numerically the
self consistency equations (33)  for each structure separately and compare the free energies of the solutions. 

Interestingly, the theory predicts that the nematic splay-bend 
phase can be stabilized directly from $N$ for not too thick molecules.


\subsection{The SB molecules}

\begin{figure}[htb]
\centerline{%
\includegraphics[width=0.6\columnwidth]{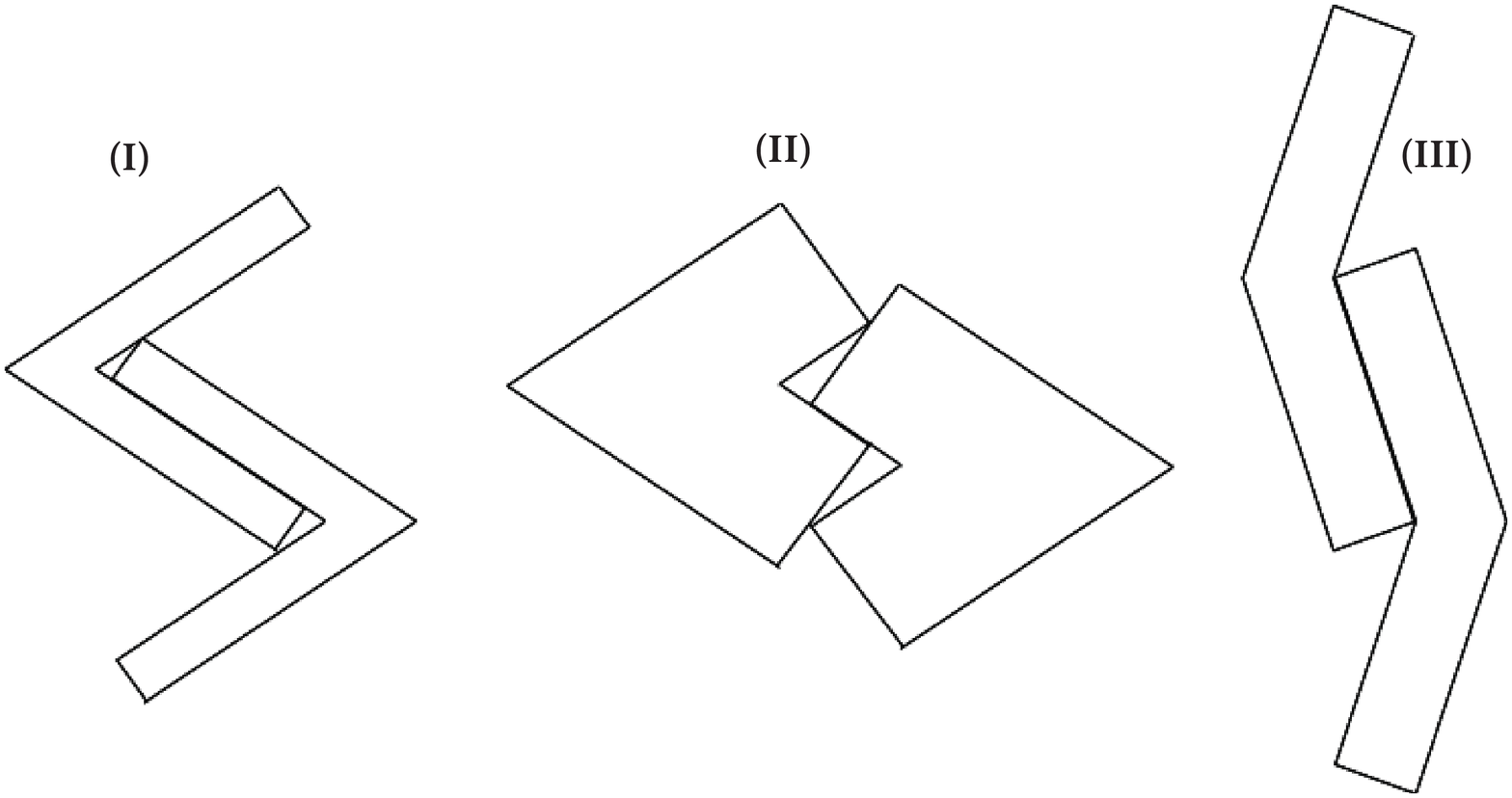}}
\caption{Three cases of antiparallel arrangements for SB molecules. They lead to  three different Fourier transforms of excluded slice given in Appendix A.}
\label{Fig10_fig:cases}
\end{figure}

In the case of the SB molecules the calculation of the Fourier transforms for the excluded slice becomes more involved. Firstly, the particle  width $\delta$ has to be smaller than $\tan(\psi)$ due to geometrical constraints. Secondly, three cases with different antiparallel arrangements (I, II, III)  of the molecules (see Fig.~\ref{Fig10_fig:cases}) have to be considered separately. The corresponding $\alpha$ and $\beta$  functions entering the bifurcation equation, like those in Eqs.~(48), should now be replaced by  $\alpha_i$  and  $\beta_i$ ($i=I,II,III$), respectively.
The first two cases appear when the opening angle $2 \psi$ is smaller than $\pi /2$. The first one of these two
occurs also when arms are thin, namely when their width is smaller than $1/[\cot(\psi) + \csc(2 \psi)]$. The corresponding normalized $\alpha_i$ and $\beta_i$ functions for $i=$ I, II, III are 
given in 
Appendix A. 

\begin{figure}[htb]
\includegraphics[width=0.45\columnwidth]{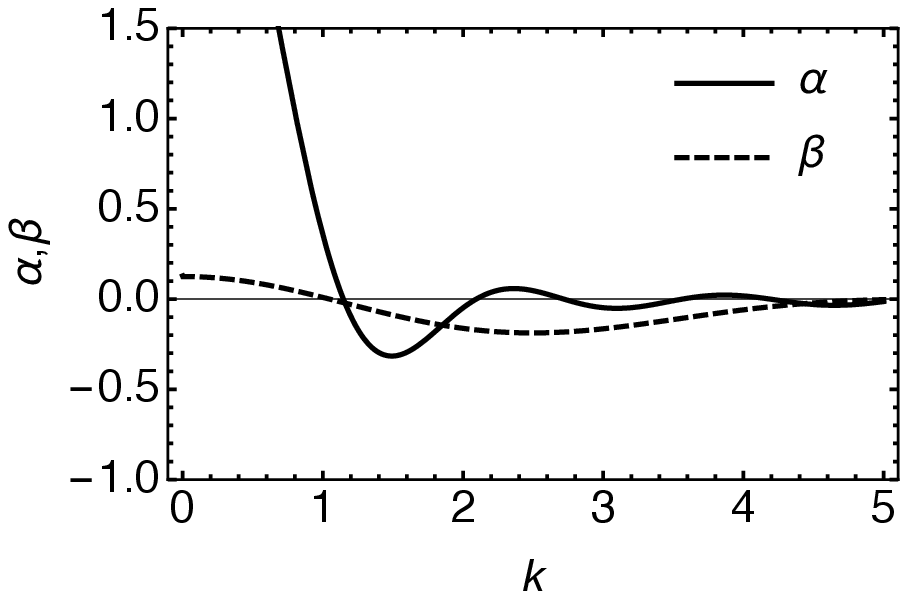}
\hspace{0.05\columnwidth}
\includegraphics[width=0.45\columnwidth]{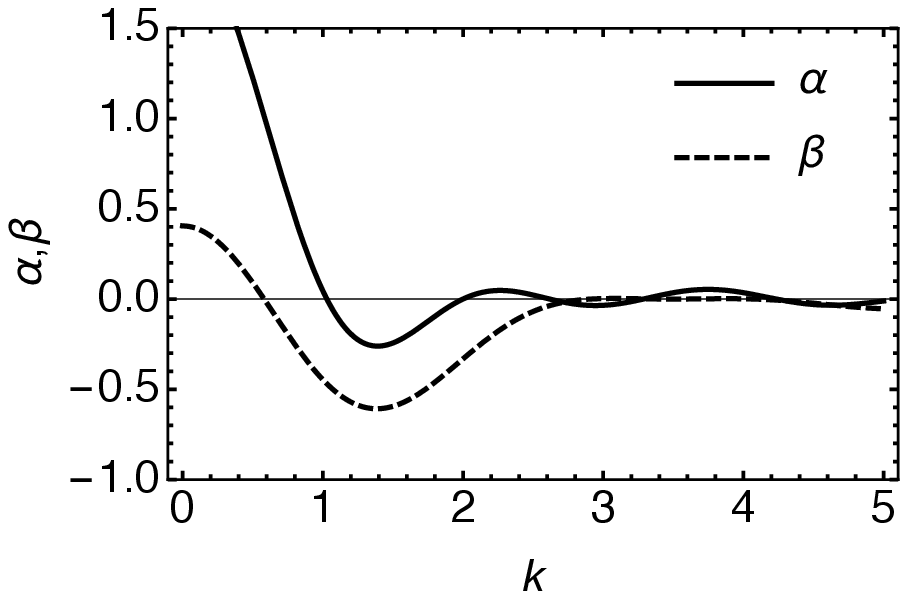}
\caption{Different relations between absolute minimum of $\alpha$ and $\beta$. Left diagram corresponds to $\psi = \pi / 4$ and  $\delta = 0.5$ and bifurcation to smectic A phase. Right diagram corresponds to $\psi = \pi / 4$ and  $\delta = 0.1$, where bifurcation is to antiferroelectric smectic phase.}
\label{Fig11_fig:ab}
\end{figure}

\begin{figure}[htb]
\centerline{%
\includegraphics[width=0.6\columnwidth]{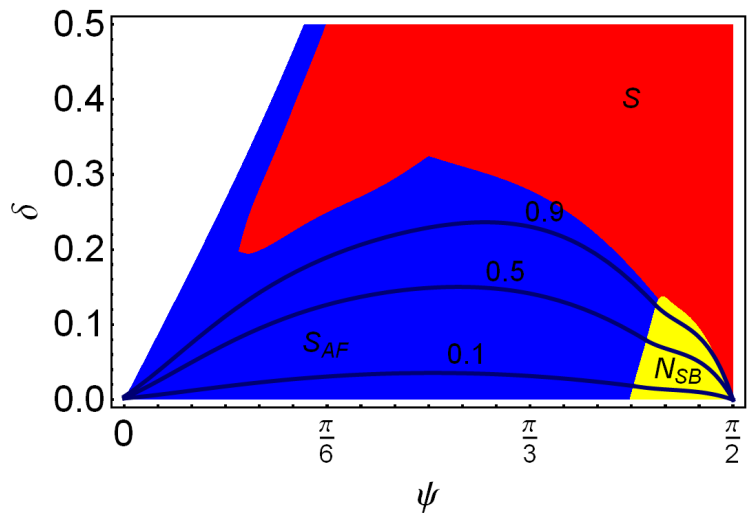}}
\caption{(Colour online) Bifurcation diagram for SB boomerangs. The blue region corresponds to the cases where nematic ($N$)-antiferroelectric smectic A ($S_{AF}$)  bifurcation takes place and the red region corresponds to the cases where the nematic ($N$)-smectic A ($S_A$) or nematic ($N$)-smectic F ($S_F$) bifurcation undergoes. In the yellow region $N_{SB}$ bifurcates from $N$. The black lines are the density limits with the  packing fraction $\eta$  given in the legend.}
\label{Fig12_fig:bif}
\end{figure}

Examining the positions of relative minima for $\alpha_i$ and $\beta_i$  (see Fig.~(\ref{Fig11_fig:ab})) one 
observes bifurcations to different phases as shown in Fig.~(\ref{Fig12_fig:bif}).
More specifically, the first panel of Fig.~(\ref{Fig11_fig:ab}) illustrates the case where the bifurcation to $S_A$
or  a transition to 
$S_F$ takes place, which is connected with the coefficient $\alpha$ having  a global negative minimum  deeper than
that  of $\beta$. The next panel shows the opposite case, {\textit{i.e.}} when the (negative) global minimum of $\beta$
is deeper than that of $\alpha$, hence the 
bifurcating phase will be $S_{AF}$. When the (negative) global minima  are about the same depth 
we can expect an incommensurate smectic phase to become absolutely stable.
Within our formalism this case can be studied by taking a commensurate approximation, where both minima
are approximated by an appropriate choice of $k,k'$ and $M,M'$.

Similar to  the HB case  the coefficient $\beta$ at $k_m=0$ is always positive, thus the polar nematic phase
cannot appear here either.  The complete bifurcation diagram is presented in Fig.~(\ref{Fig12_fig:bif}).
The blue region in this diagram corresponds to the apex angle and arm thickness  where the bifurcation
from $N$  to $S_{AF}$ takes place. The red area ($S$) again corresponds to $N-S_A$ bifurcation or $N-S_F$  
transition and in the yellow area the bifurcation is dominated by $N_{SB}$. The black lines are the guide lines at 
which the packing fractions are given as in the legend.

Finally, we would like to add that we have checked a few cases for a possibility to get stable 
incommensurate smectic phases  of A type and a smectic C phase, where the director is not 
perpendicular to the layer normal, but found no one more stable than the structures identified in Figs.~(\ref{Fig09_bifuPawel}, \ref{Fig12_fig:bif}).

\subsection{Nematic splay-bend}

\begin{figure}[htb]
\centerline{%
\includegraphics[width=0.4\columnwidth]{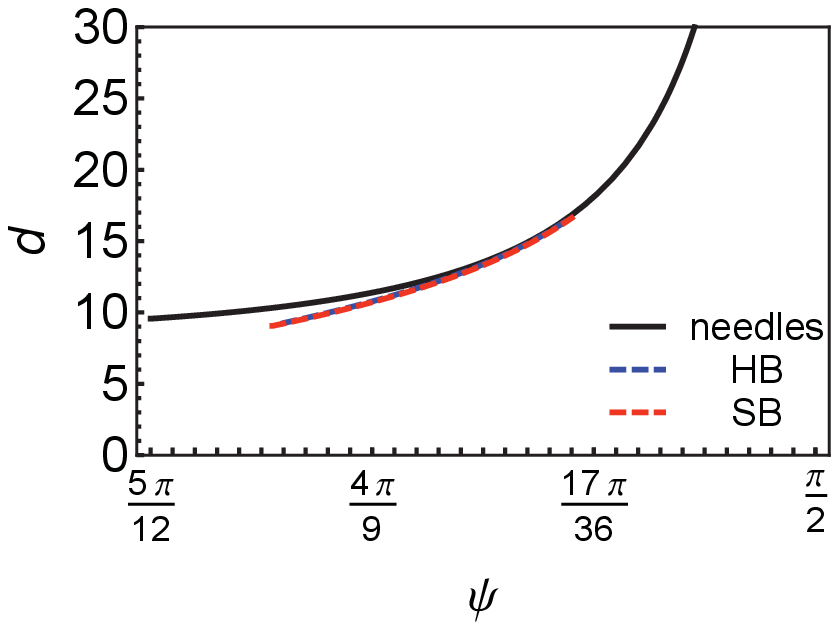}
\hspace{0.05\columnwidth}
\includegraphics[width=0.4\columnwidth]{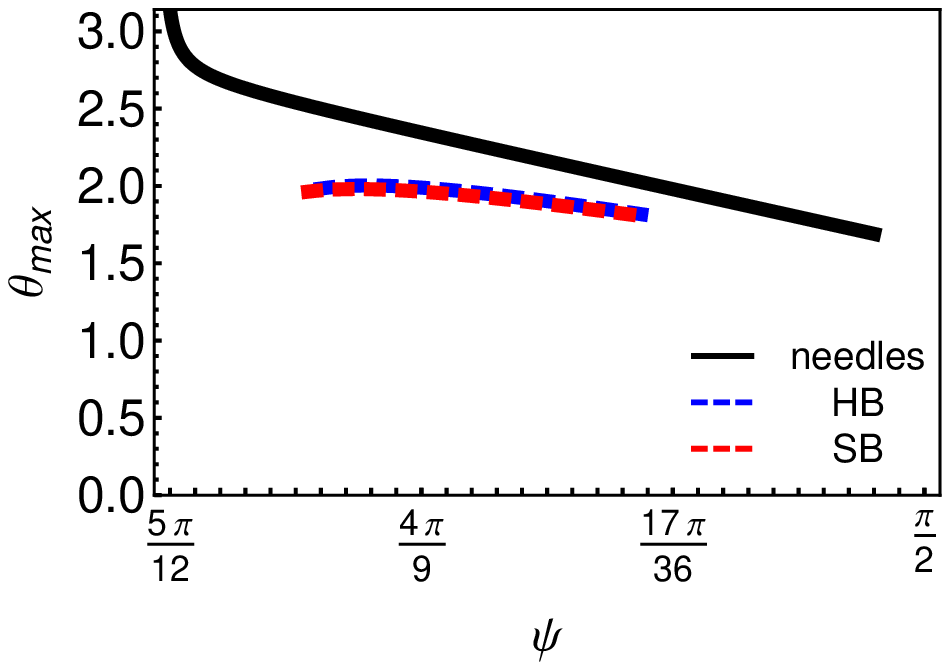}}
\caption{(Colour online) Behaviour of  period $d$ and $\theta_{max}$ as function of $\psi$ for $N_{SB}$ calculated at bifurcation from the nematic phase. Blue and red lines correspond to  HB and SB particles of width $\delta=0.1$.}
\label{Fig20_fig:theta}
\end{figure}

Both, HB and SB bifurcation diagrams contain previously described nematic splay-bend regions, 
but the range of stability of  
 $N_{SB}$ is narrow, starting from $\psi$ around $5\pi/12$. Another limiting factor is the width $\delta$ of 
a particle, which cannot be greater than $0.14$. In Fig.~(\ref{Fig20_fig:theta}) further characteristics of 
$N_
{SB}$ 
are shown.  In particular, please note that the period $d$  of  $N_{SB}$ increases with increasing opening 
angle, 
but  width of a  particle has no  significant effect on $d$. It only reduces the range of $\psi$  at which the $N_{SB}$ 
phase 
can occur. However,  
$\theta_{max}$ 
can be drastically altered by the width of a particle. For  hard needles 
of  $\psi=5\pi/12$ it approaches its maximal value  of $\pi$, which means that molecules perform a full half-
turn on the path of length  $d/2$,   but as  thickness of   particles increases
$\theta_{max}$   gets smaller.  The same effect is observed when
the opening angle of molecules increases. It causes    $\theta_{max}$  to decrease towards $\pi/2$, which means that
the  tilt angle of the director with respect to the $x$-axis varies between  $\pi/4$  and  $-\pi/4$.

Please remember that we parametrize our results using convention adopted  for bent-core needles [25]. As already mentioned before this parametrization  is  singular in the needle limit  $\psi=0, \pi/2$ due to the  factor $l^2 sin(2 \psi)$  in Eq. (2), where  bent-core molecules of zeroth thickness become reduced to a line. Therefore  any polar order that may occur  for  $\psi=\pi/2$   in Figs.~(\ref{Fig09_bifuPawel}, \ref{Fig12_fig:bif}) is only asymptotically stable, for $\eta \to \infty$.

\subsection{Exemplary results of full minimization }

Here the free energy of different  phases is calculated for
exemplary molecular shapes to identify  the stable phases as function of packing fraction.
It turns out that for majority of cases the calculations involving terms up to $n=m=4$
in (33) and (34) give excellent quantitative predictions for the equilibrium structures.
The obtained  results are consistent with the phase diagram maps,
Figs.~(\ref{Fig09_bifuPawel}, \ref{Fig12_fig:bif}),  in apex angle- arm's width plane.
Here  we  concentrate on the most common  $S_{AF}$ phase and clarify the issue of previously mentioned $S_F$ phase.

\begin{figure}[htb]
\begin{center}
\includegraphics[width=0.45\columnwidth]{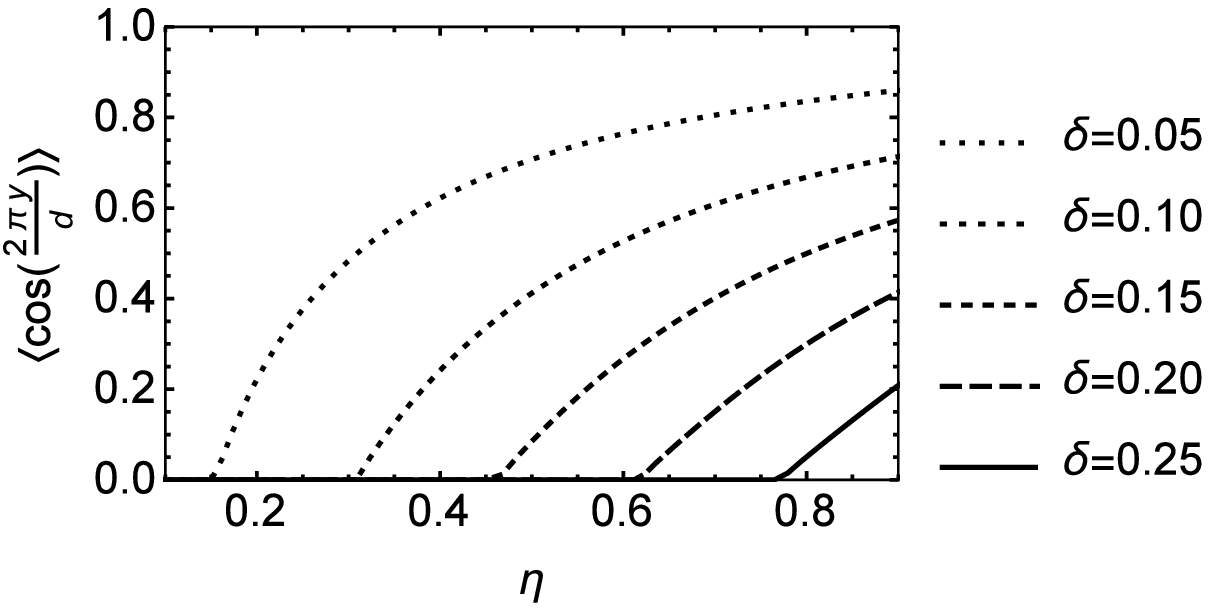}
\hspace{0.05\columnwidth}
\includegraphics[width=0.45\columnwidth]{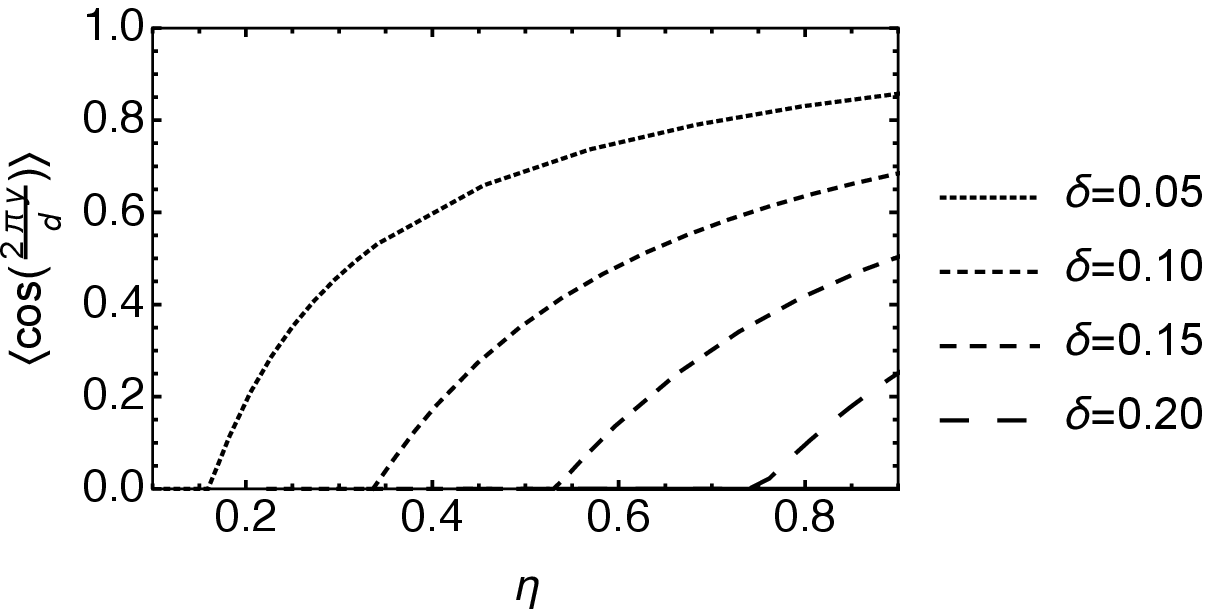}
\end{center}
\caption{{\protect\small {Typical behaviour of equilibrium order parameter $\langle c \rangle$ for the cases with
stable $S_{AF}$ phase,  obtained for different
packing fractions $\eta$ and the apex angle $2\psi=2\pi/3$.
The panel  on the left is for the HB molecules
while the panel on the right is for the SB molecules.} }}
\label{Fig13_orderparamAF}
\end{figure}

\begin{figure}[htb]
\begin{center}
\includegraphics[width=0.45\columnwidth]{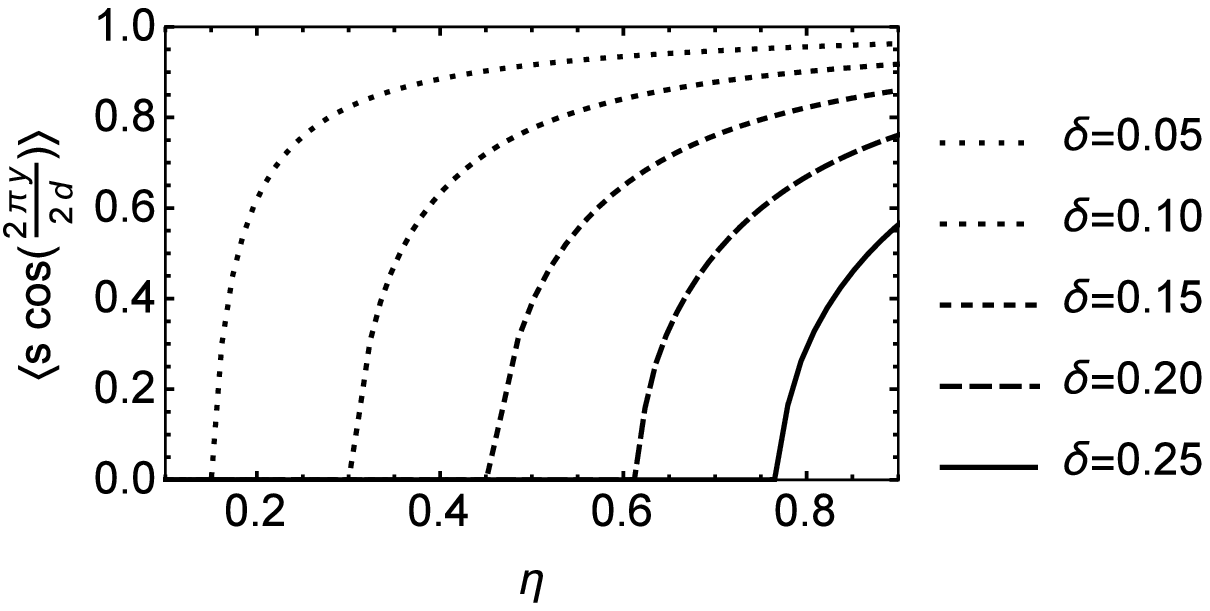}
\hspace{0.05\columnwidth}
\includegraphics[width=0.45\columnwidth]{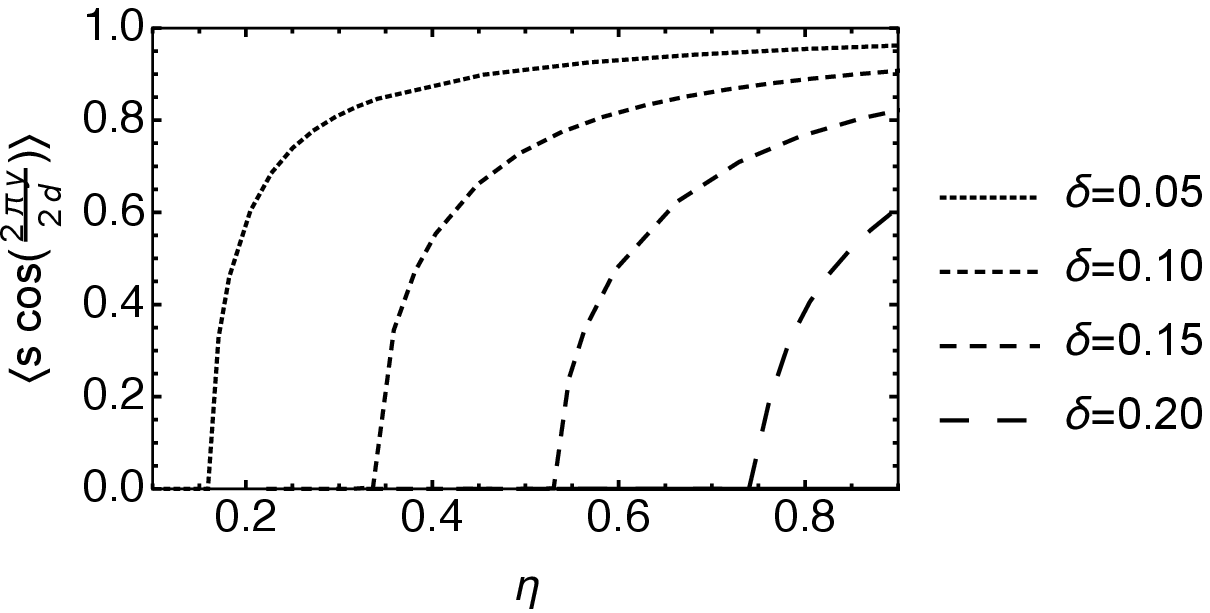}
\end{center}
\caption{{\protect\small {Typical behaviour of equilibrium order parameter $\langle sc \rangle$ for the cases with stable $S_{AF}$ phase, obtained for different packing fractions $\eta$ and the apex angle $2\psi=2\pi/3$.
The panel on the left is for the HB molecules while
the panel on the right is for the SB molecules.} }}
\label{Fig14_orderparamAFsc}
\end{figure}

We start with the case of  stable $S_{AF}$, represented by the blue region
in Figs.~(\ref{Fig09_bifuPawel}, \ref{Fig12_fig:bif}).   In Figs.~(\ref{Fig13_orderparamAF}, \ref{Fig14_orderparamAFsc}) we compare the equilibrium values of leading order parameters
for this structure as function of packing fraction
and  arm's width for $\psi=\pi/3$.
It turns out that the change of the arm endings strongly influences the behaviour
of the order parameters, especially when the thickness of the arms increases.
As expected, for thin arms, where for instance $\delta=0.05$, the profiles of the order parameters
(and the bifurcation points) are very similar.
 For larger values of $\delta$ the bifurcation point for the
 SB boomerangs shifts toward higher packing fractions.
 For $\delta=0.25 $ one does not observe the stable $S_{AF}$  phase,
 whereas for the HB boomerangs this structure is still attainable.
\begin{figure}[htb]
\begin{center}
\includegraphics[width=0.45\columnwidth]{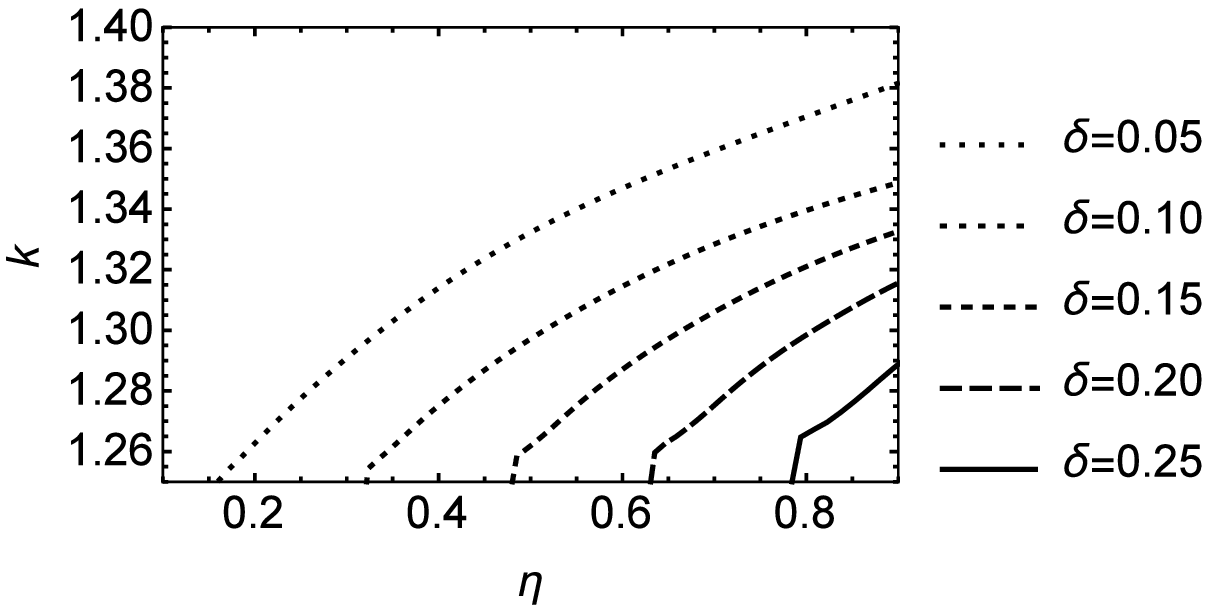}
\hspace{0.05\columnwidth}
\includegraphics[width=0.45\columnwidth]{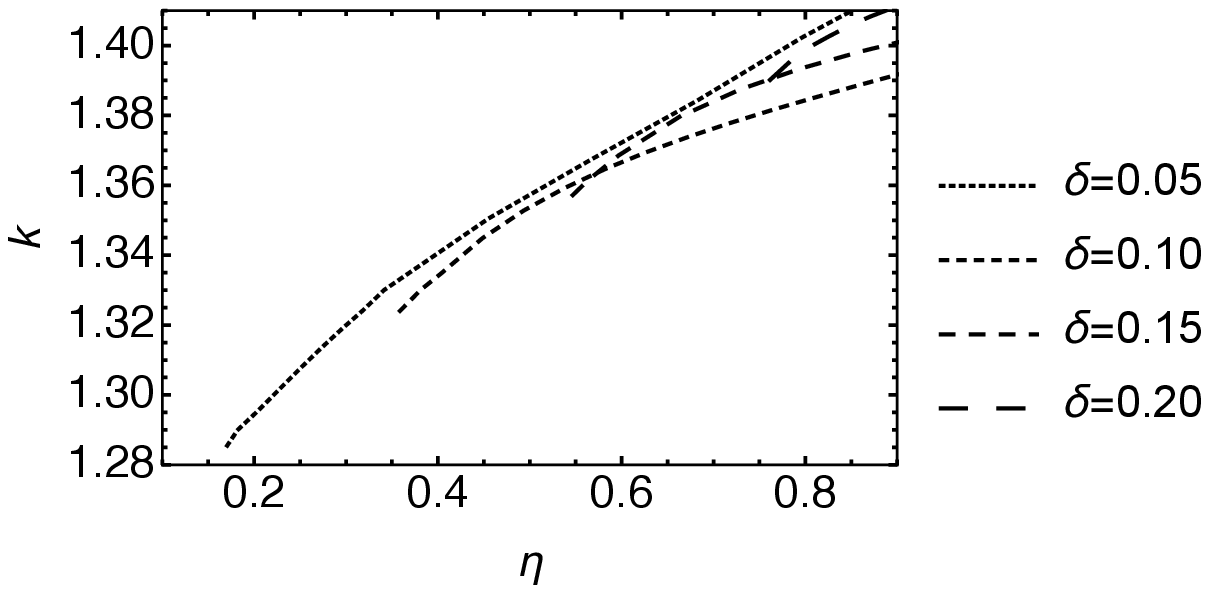}
\end{center}
\caption{{\protect\small {Equilibrium wave vector $k$ ($k'=k/2$) in stable $S_{AF}$ phase, obtained for different packing fractions and the apex angle $2\psi=2\pi/3$ for the HB molecules (left) and for the SB molecules (right). The layer thickness is proportional to inverse of $k$.} }}
\label{Fig15_Kappas3}
\end{figure}
 \begin{figure}[htb]
\begin{center}
\includegraphics[width=0.45\columnwidth]{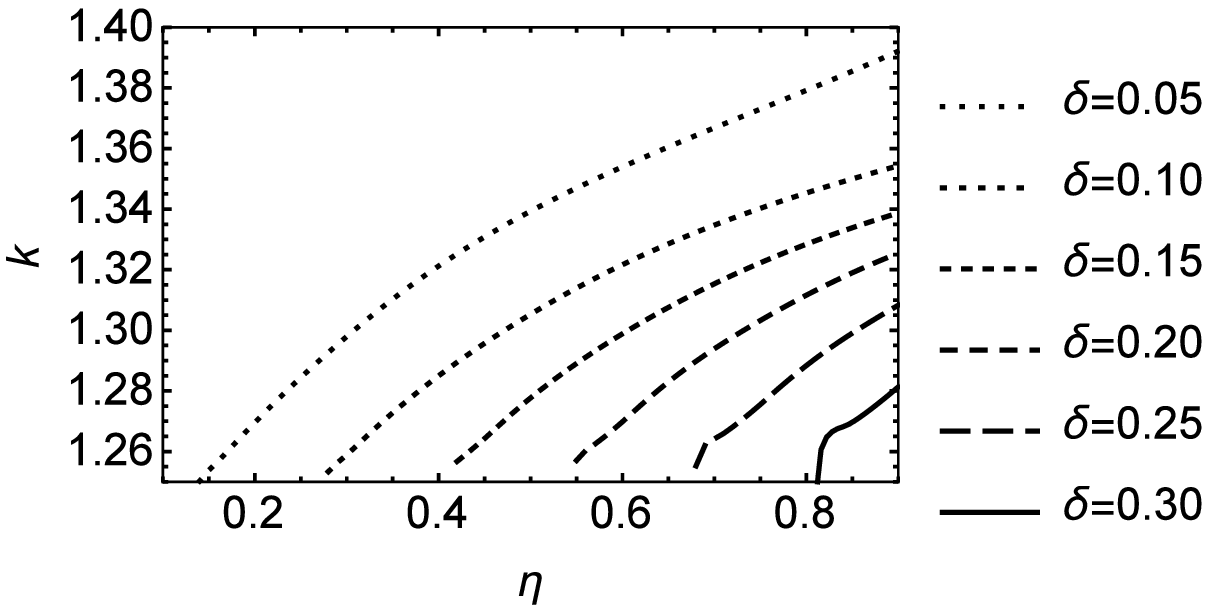}
\hspace{0.05\columnwidth}
\includegraphics[width=0.45\columnwidth]{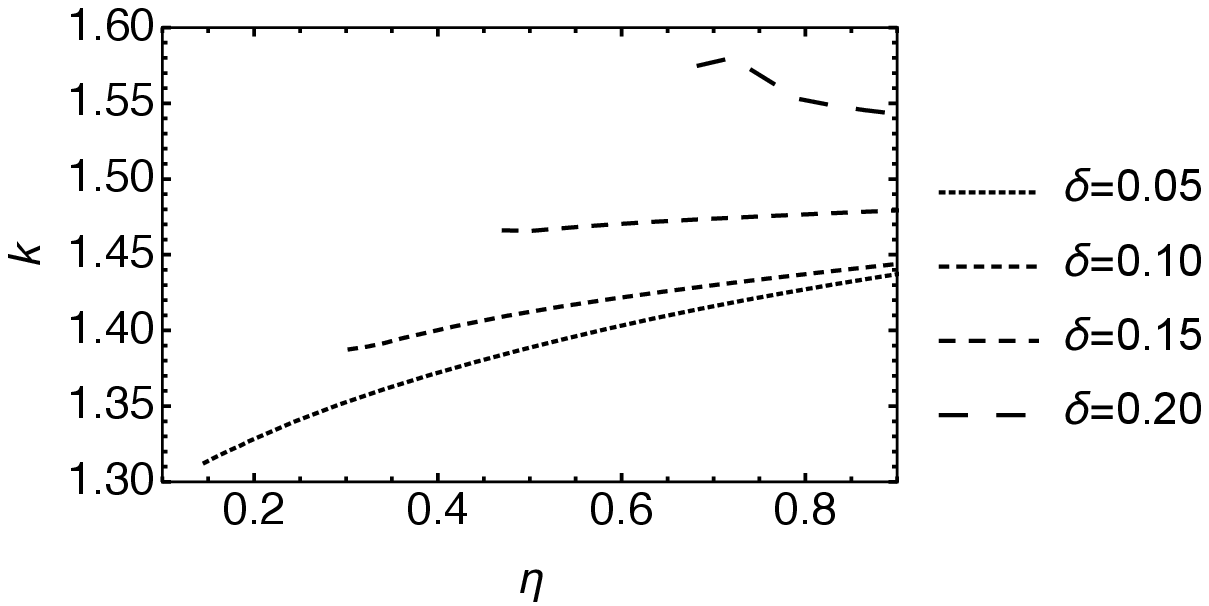}
\end{center}
\caption{{\protect\small {Equilibrium wave vector $k$ ($k'=k/2$) in stable  $S_{AF}$ phase, obtained for different packing fractions and the apex angle of $2\psi=2\pi/4$ for the HB molecules (left) and for the SB molecules (right). The layer thickness is proportional to  inverse of $k$.
} }}
\label{Fig16_Kappas4}
\end{figure}

In Figs.~(\ref{Fig15_Kappas3}) and (\ref{Fig16_Kappas4}) the equilibrium wave vector $k$ of the $S_{AF}$ 
phase, obtained for different packing fractions   $\eta$
and the apex angle of $2\psi=2\pi/3$ and  $2\psi=2\pi/4$, respectively, is presented
for HB and SB molecules. In case of HB molecules the  $k$ vector increases with the packing fraction, which  means diminishing of the layer thickness. For the SB molecules  the wave vector $k$ can show  different behaviour. In the case  of $2\psi=2\pi/3$,  for thicker arms ($\delta>0.15$), the wave vector
diminishes with the packing fraction and hence  the layer thickness increases.

\begin{figure}[htb]
\begin{center}
\includegraphics[width=0.4\columnwidth]{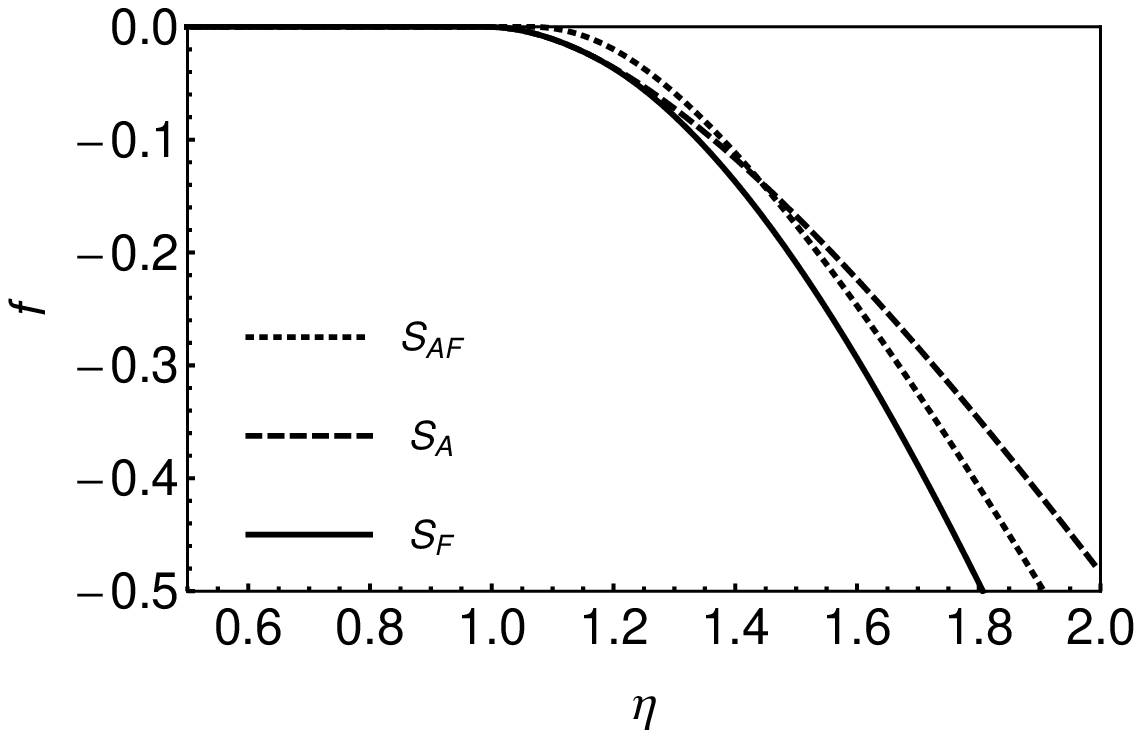}
\hspace{0.05\columnwidth}
\includegraphics[width=0.4\columnwidth]{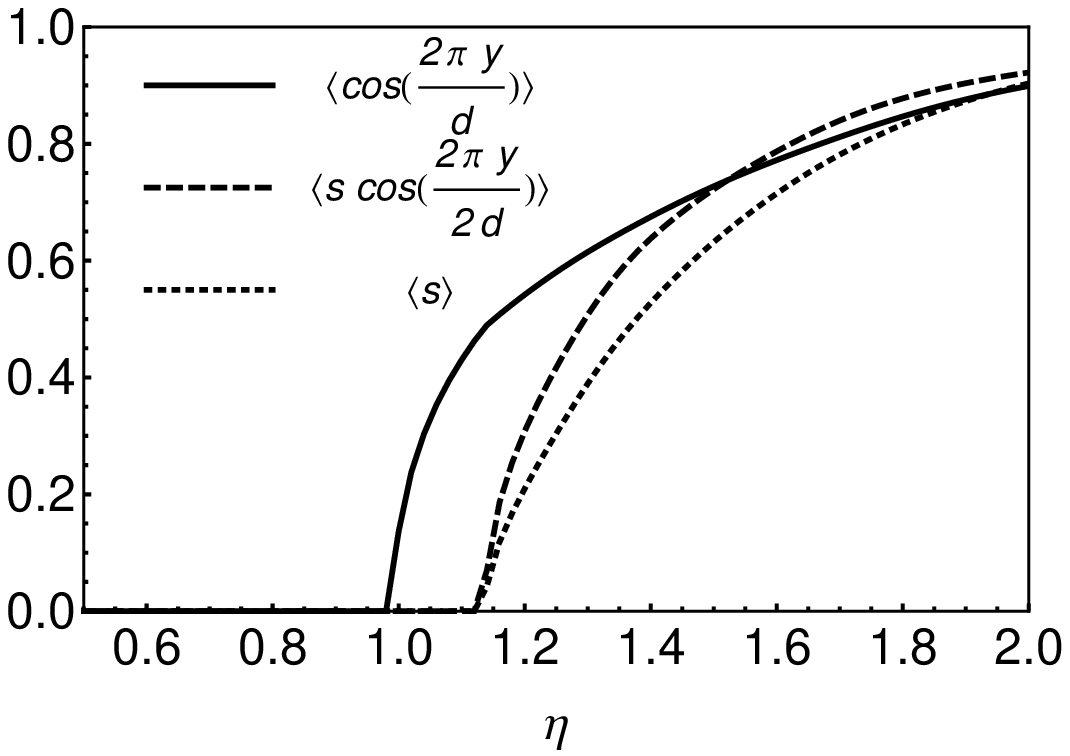}
\end{center}
\caption{{\protect\small { Dependence of free energies  on  packing fraction $\eta$  for three structures: $S_{AF}$,  $S_A$ and  $S_F$  of  HB system  (left panel) and equilibrium, leading order parameters for stable $S_A$ and  $S_F$  (right panel). Molecular parameters are $\delta=0.4$ and $\psi=\frac{\pi}{4}$. Nematic phase ($f=0$) is stable for $\eta<1$ while $S_A$ wins for $\eta>1$.  $S_F$ becomes more stable than $S_A$ for  $\eta>1.15$. Note that lamellar phases with polarized layers are close-packed ground states for  HB systems. } }}
\label{Fig17_free_energy04}
\end{figure}

\begin{figure}[htb]
\begin{center}
\includegraphics[width=0.4\columnwidth]{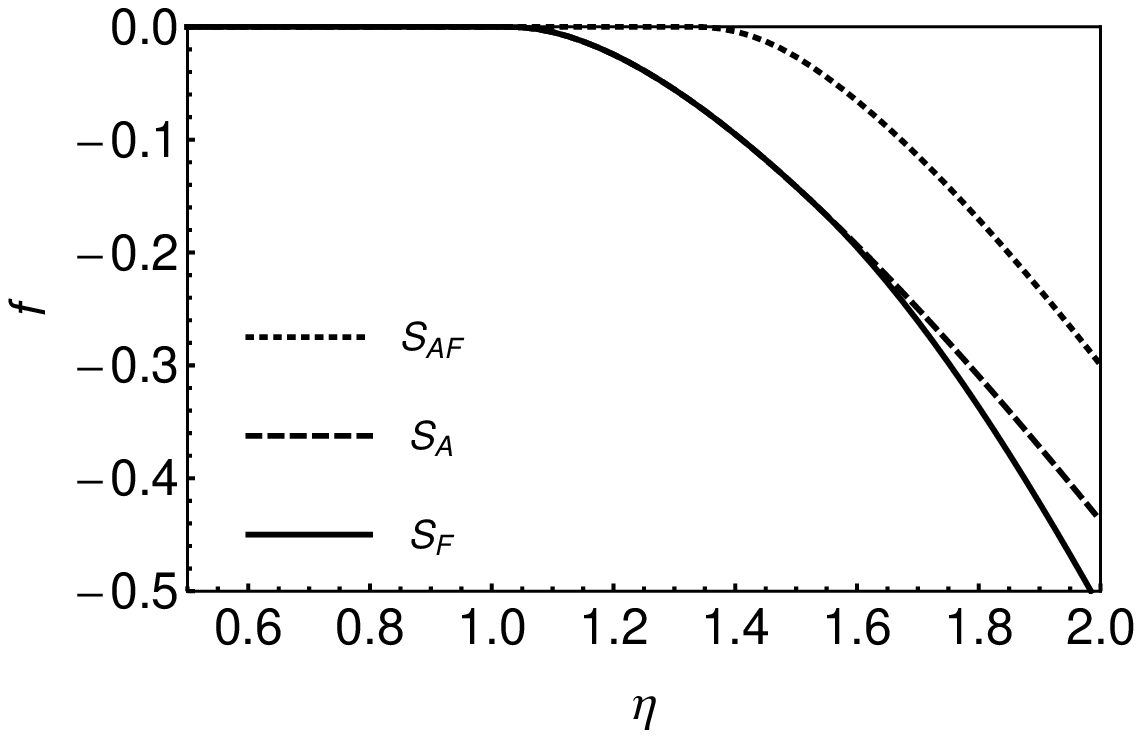}
\hspace{0.05\columnwidth}
\includegraphics[width=0.4\columnwidth]{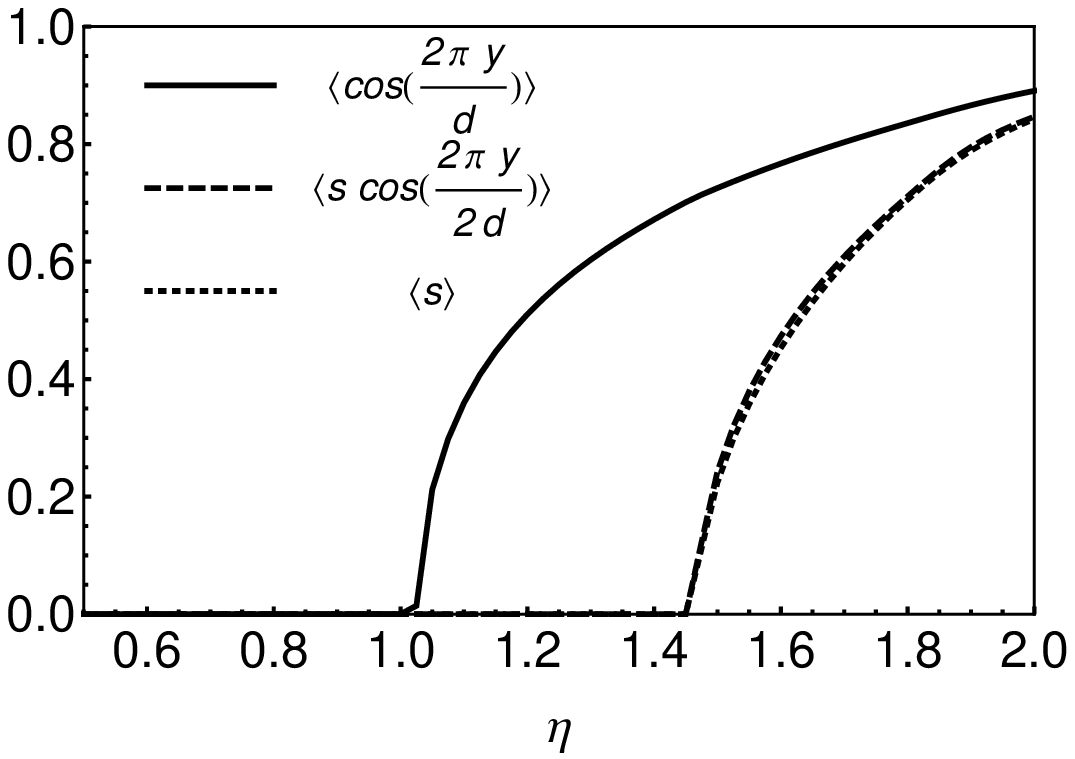}
\end{center}
\caption{{\protect\small {
 Dependence of free energies  on  packing fraction $\eta$  for three structures: $S_{AF}$,  $S_A$ and  $S_F$  of  HB system  (left panel) and equilibrium, leading order parameters for stable $S_A$ and  $S_F$   (right panel). Molecular parameters are $\delta=0.5$ and $\psi=\frac{\pi}{4}$.  Nematic phase ($f=0$) is stable for $\eta<1$ while  $S_A$  wins for $\eta>1$.   $S_F$ 
becomes more stable than $S_A$  for  $\eta>1.45$. Note that lamellar phases with polarized layers are close-packed ground states for  HB systems.
} }}
\label{Fig17_free_energy05}
\end{figure}

In order to determine the sequence of phase transitions and establish relations between them in the red ($S$) regions of the bifurcation diagrams we compared the free energies for the reference structures using the first terms in (33) and (34), and then calculated the order parameters up to $n=m=4$. The results showing stable $S_A$  and $S_F$  phases are shown in Figs.~(\ref{Fig17_free_energy04}) and (\ref{Fig17_free_energy05}) for two HB systems: near  ($\delta=0.4$, $\psi=\pi/4$) and  far  ($\delta=0.5$, $\psi=\pi/4$)  from the blue region. For the cases studied the first transition is  $N-S_{A}$ followed by $S_{A}-S_{F}$ for larger $\eta$s. The distance between both transitions grows with the width of a particle. Note that the structures became stable at high  packing fractions $\eta$.

%

\subsection{Exemplary MC simulation}

\begin{figure}[htb]
\begin{center}
\includegraphics[width=0.45\columnwidth]{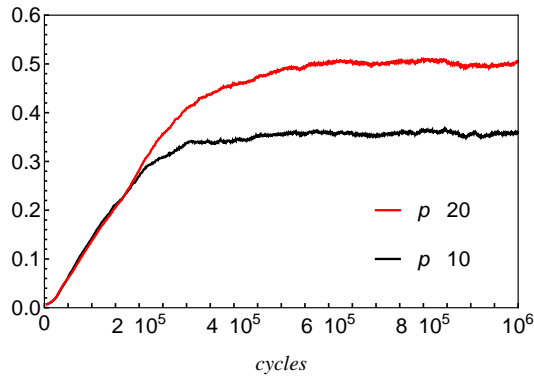}
\end{center}
\caption{{\protect\small {(Colour online) Exemplary dependence of packing fraction on MC cycle number. Simulations were performed for $N=500$ SB particles, for $\psi=\frac{5}{12}\pi$ and $\delta=0.1$. Pressures are given in the legend. Note that equilibration of the system  is achieved after about  $4 \cdot 10^5$ cycles for $p=10$ and $6 \cdot 10^5$ cycles for $p=20$.}
}}
\label{fig:eta_cycles}
\end{figure}

In order to check whether the approximation of ideal nematic order gives correct 
qualitative predictions for our models we carried out exemplary, constant pressure 
MC simulations for needle like, HB and SB boomerangs. All simulations began with a 
set of $N=500$ to $N=2000$ particles of the same type, randomly oriented and placed 
inside a box with periodic boundary conditions applied. A single MC step involved 
random selection of a particle and a random translation and rotation, accepted only 
if the particle did not intersect with any others.
$N$ of such steps were considered as a single cycle. The size of the simulation box 
was dynamically adjusted to keep the constant pressure of the system. The rescaling 
of the box took place every 10 cycles and was successful with a probability
\begin{equation}
1 - \left( \frac{S_{new}}{S_{old}} \right)^N \exp \left[ - p (S_{new} - S_{old}), \right]
\end{equation}
if the particles in new positions did not intersect. Here $S_{old}$ and $S_{new}$ 
denote the surface of the box before and after rescaling, respectively, and $p$ 
is the pressure. These transformations were adjusted to keep the MC acceptance ratio 
between $0.3 - 0.5$. 
The dependence of the packing fraction $\eta$ on number of MC cycles, shown 
in Fig.~(\ref{fig:eta_cycles}), suggests that the system became equilibrated after 
about $4 \cdot 10^5$ cycles for $p=10$ and $6 \cdot 10^5$ cycles for $p=20$.

\begin{figure}[htb]
\centerline{%
\subfigure[needle boomerangs, $\psi=\pi/6$,  $\bar{\rho}=4.52$]{\includegraphics[width=0.32\columnwidth]{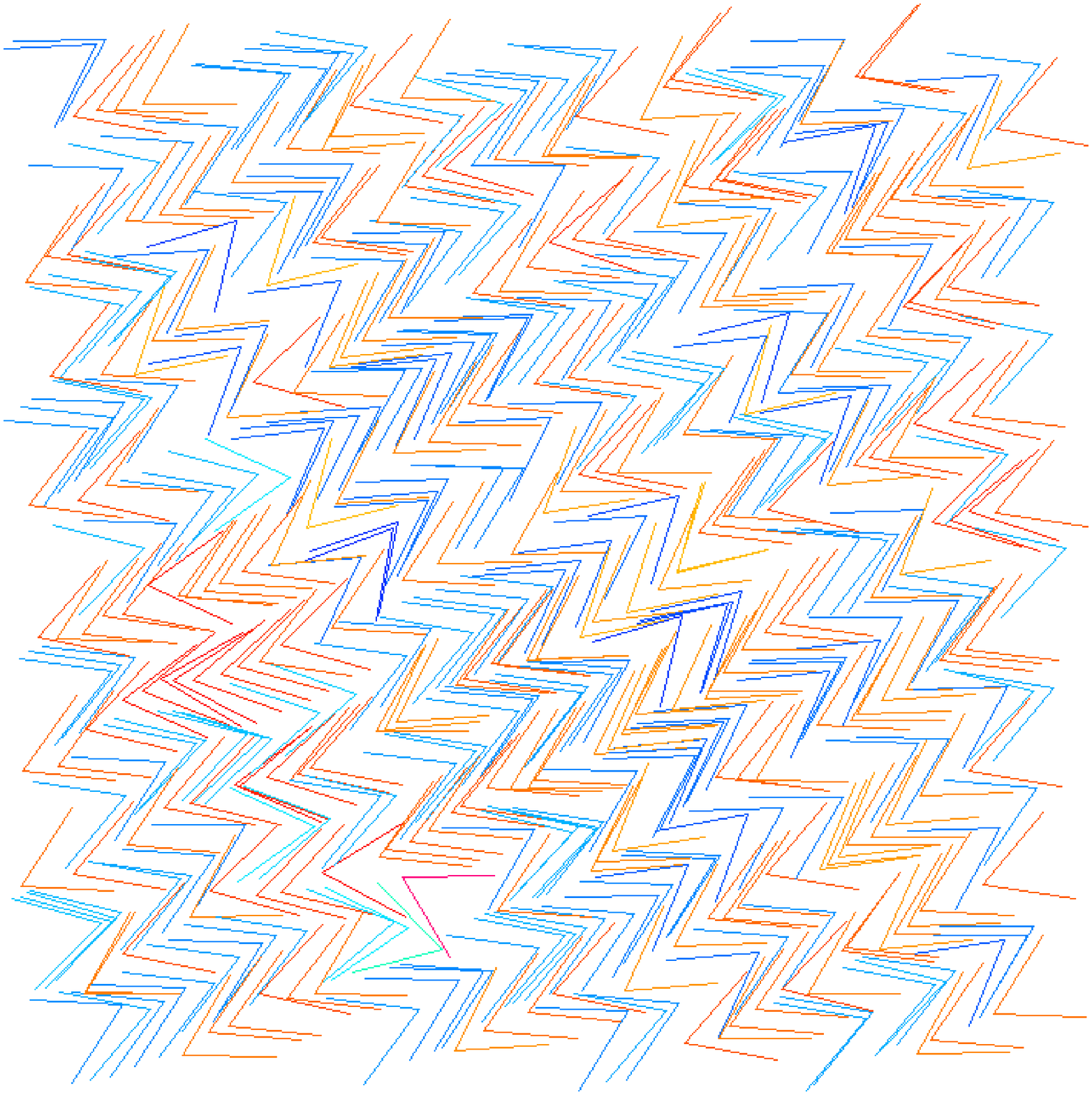}}
\hspace{0.01\columnwidth}
\subfigure[HB, $\psi=\pi/4$, $\delta=0.5$, $\eta=0.62$]{\includegraphics[width=0.3\columnwidth]{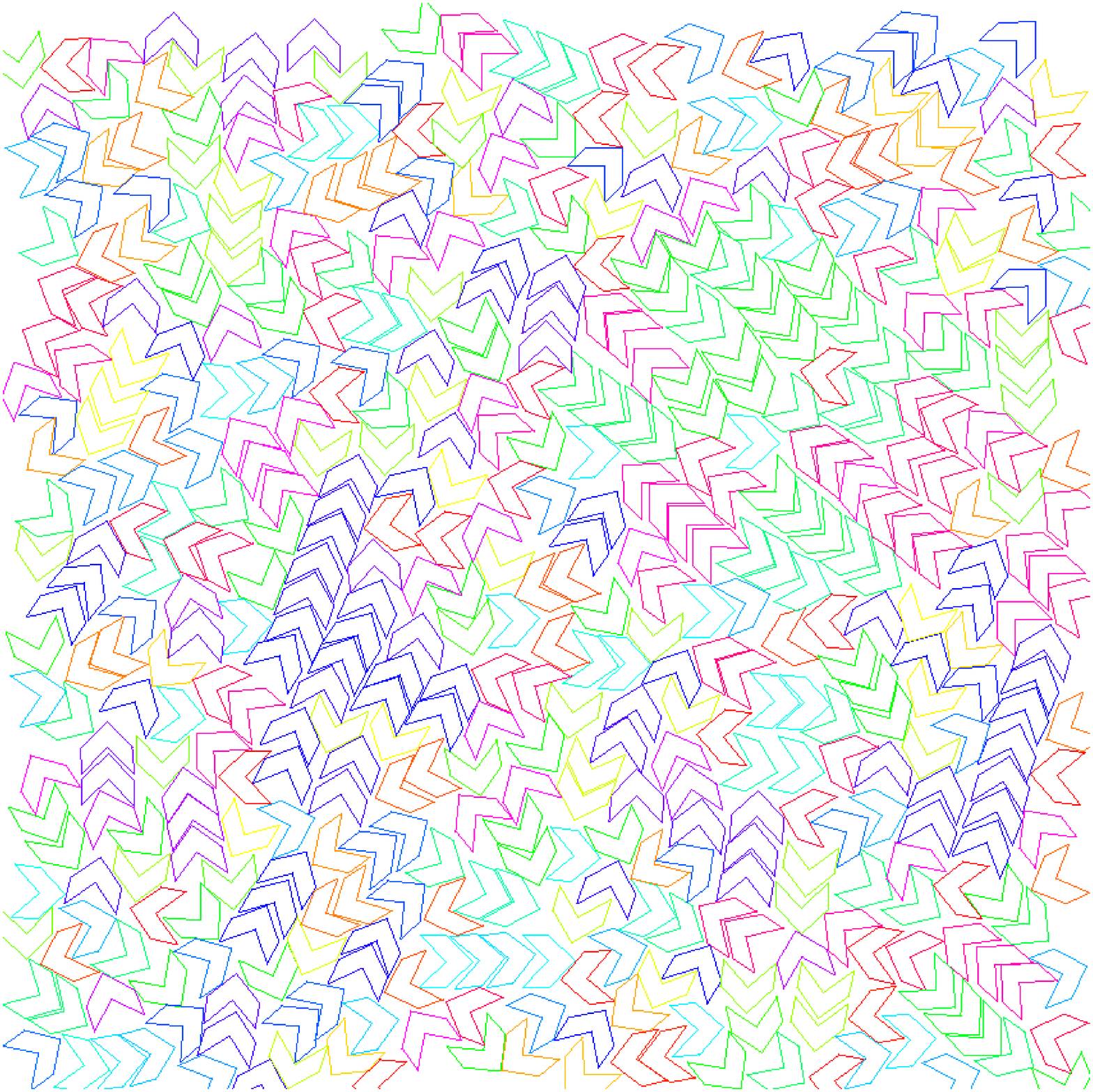}}
\hspace{0.01\columnwidth}
\subfigure[SB, $\psi=\pi/4$, $\delta=0.2$, $\eta=0.67$]{\includegraphics[width=0.3\columnwidth]{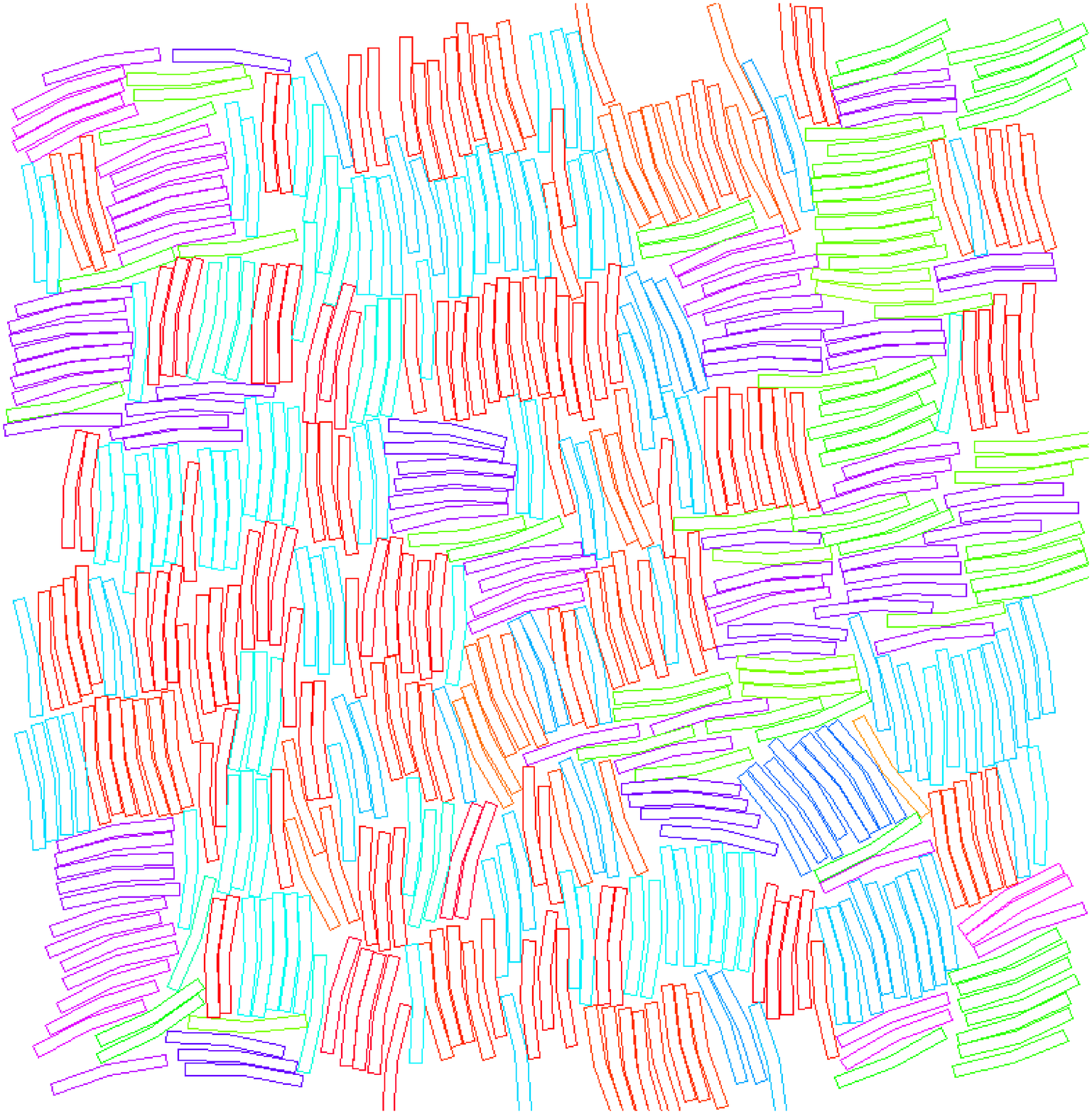}}
}
\vspace{0.01\columnwidth}
\centerline{%
\subfigure[needle boomerangs, $\psi=5\pi/12$, $\bar{\rho}=5.1$]{\includegraphics[width=0.3\columnwidth]{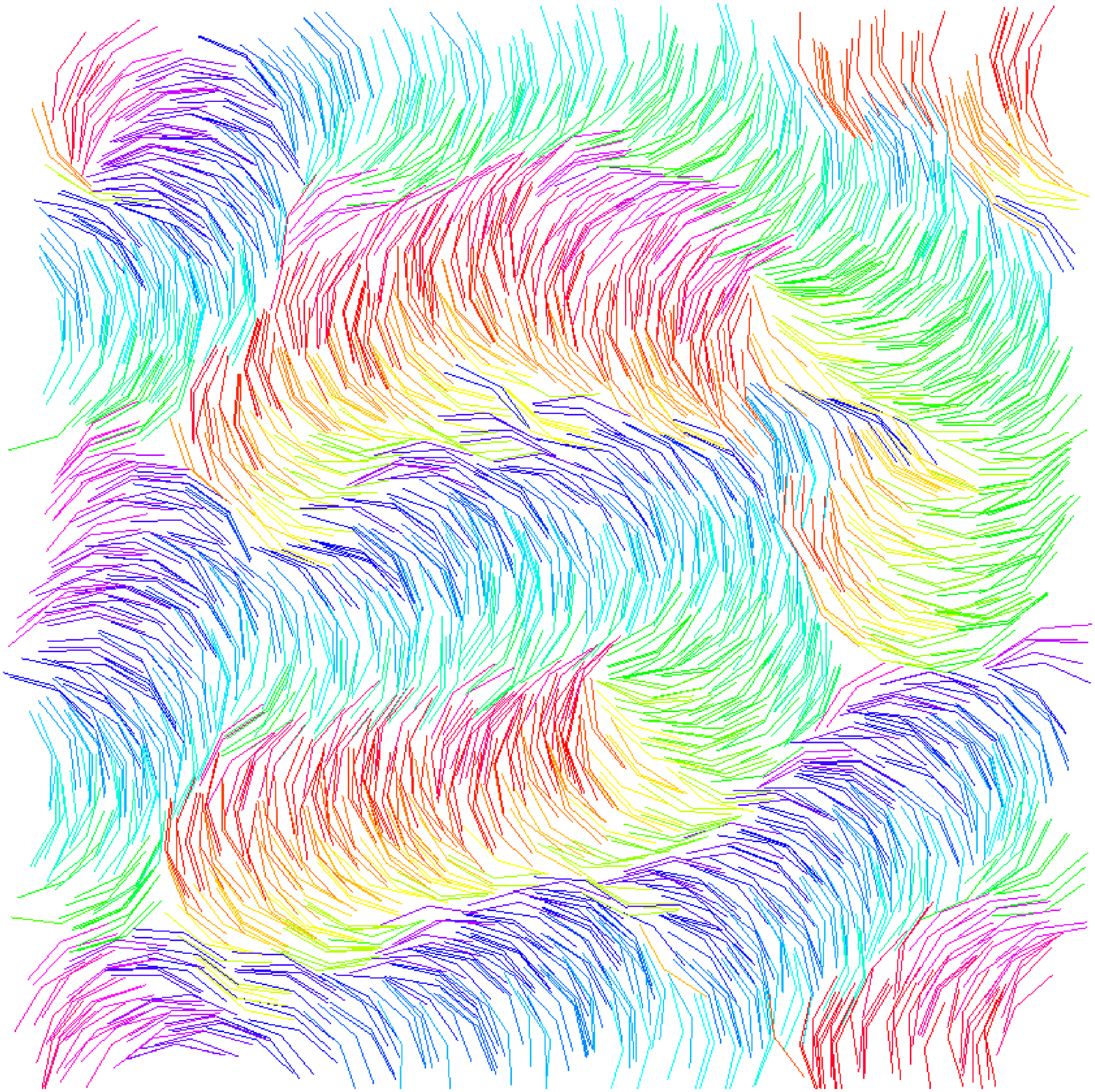}}
\hspace{0.01\columnwidth}
\subfigure[HB, $\psi=4\pi/9$, $\delta=0.1$, $\eta=0.50$]{\includegraphics[width=0.32\columnwidth]{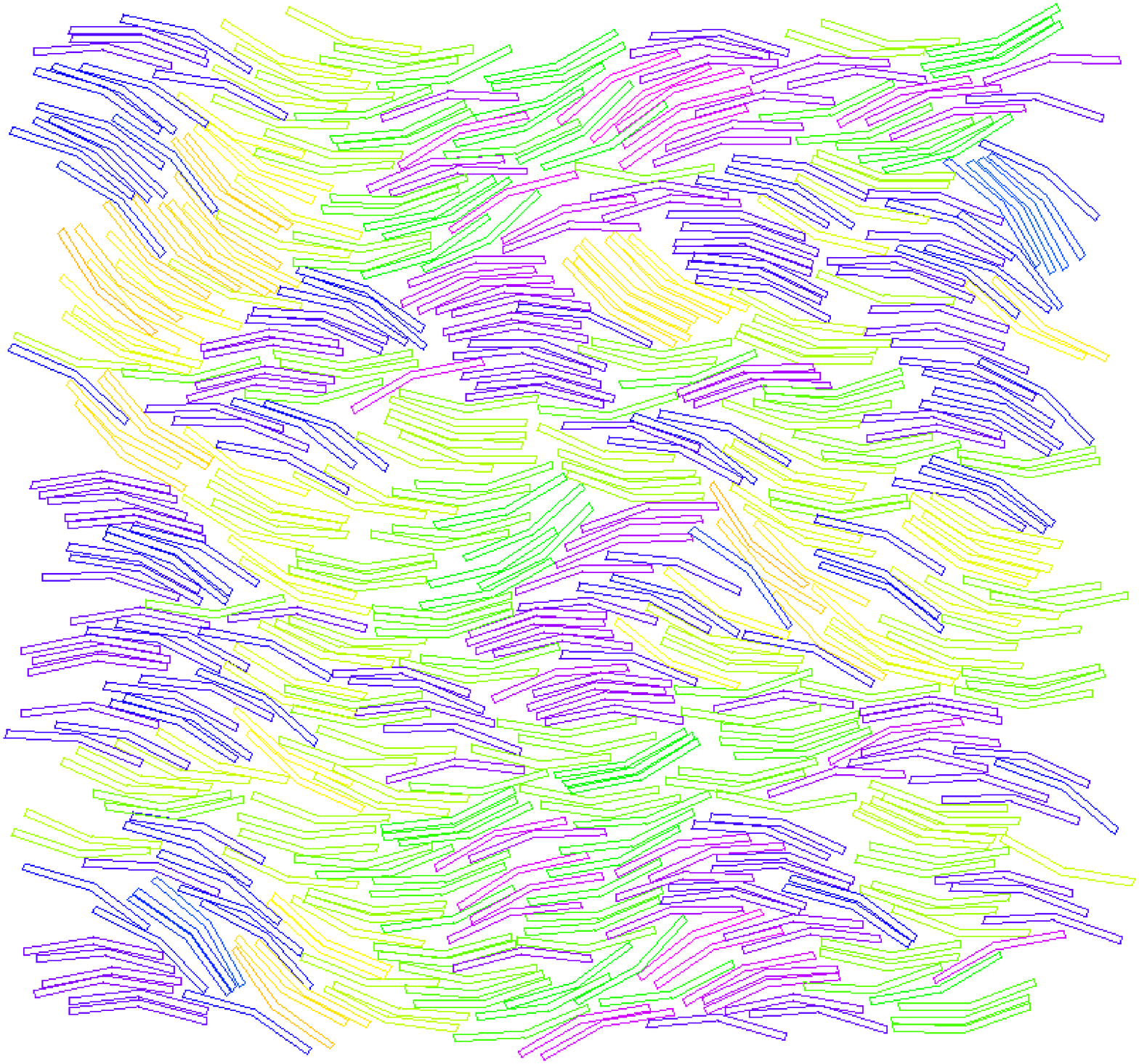}}
\hspace{0.01\columnwidth}
\subfigure[SB, $\psi=\pi/4$, $\delta=0.5$, $\eta=0.79$]{\includegraphics[width=0.32\columnwidth]{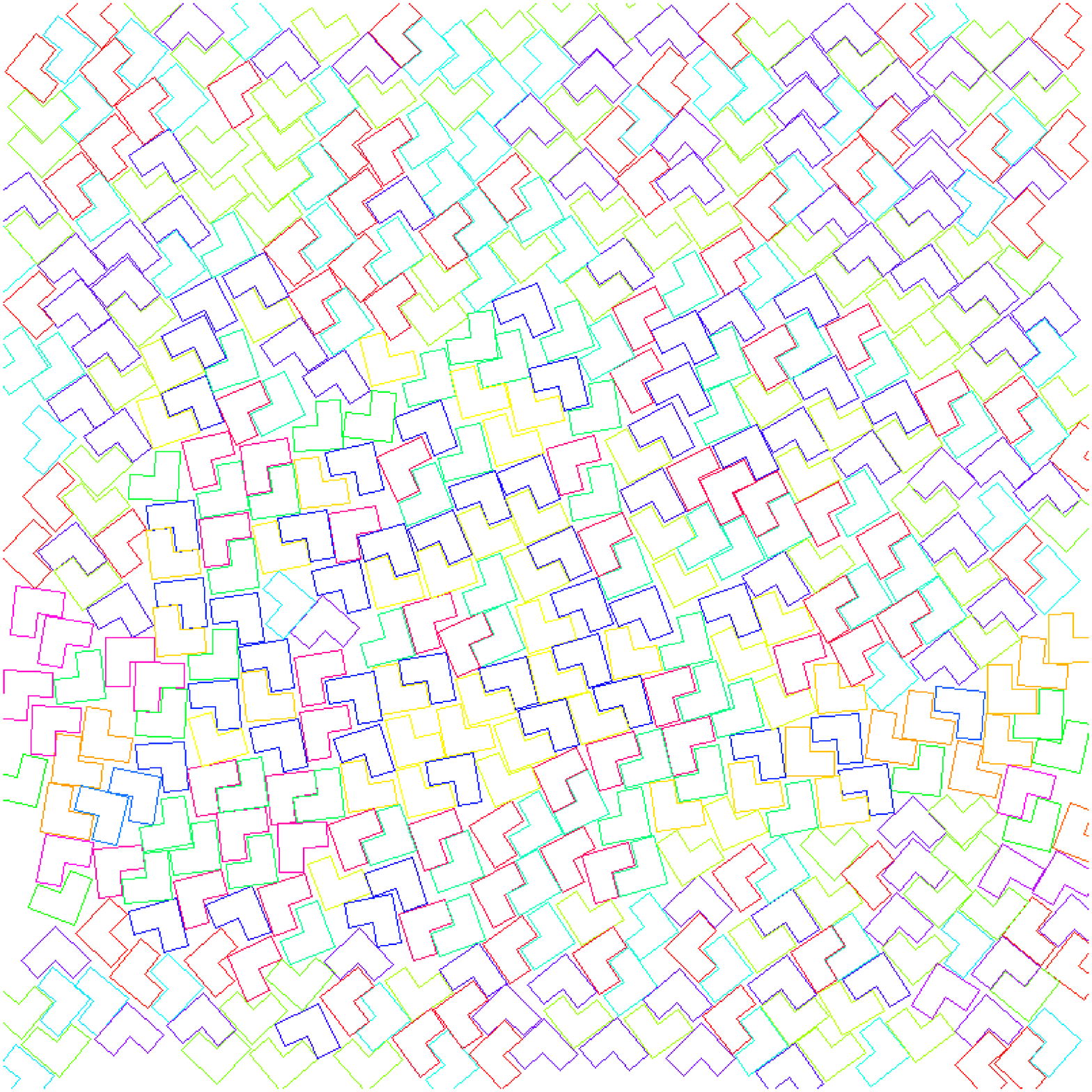}}
}
\caption{(Colour online) Exemplary snapshots from MC simulations: (a) $S_{AF}$ for needle-like boomerangs; (b) $S_{F}$ domains for boomerangs of non-zero thickness; (c) $S_A$ and $S_F$ domains; (d) nematic splay-bend for $\delta=0$  and (e) for $\delta>0$; (f) phase with local rectangular arrangement of bent-core molecules. Structures shown in panels (a) to (e) are consistent with the results of  bifurcation analysis. At high packing fractions further structures can emerge (f), that are not included in our study. Particle parameters, like width and apex angle, are given in the panels. Colour coding is used to show different molecular orientations.}
\label{fig:MCsnapshots}
\end{figure}

Exemplary snapshots taken after $10^6$  cycles are shown in Fig.~(\ref{fig:MCsnapshots}). In the panel (a) one can observe an antiferroelectric smectic phase formed by 
needle-like boomerangs. This is the type of smectic structure that can be observed in studied systems for the widest range of apex angles. For particles of nonzero thickness 
other types of smectic order can be also present, like ordinary smectic A or 
ferroelectric smectic. The panel (b), obtained for the HB boomerangs, corresponds 
to the case where domains of $S_F$  order are present. 
The panel (c) shows, on the other hand, the SB boomerangs which are almost rod-like, where two kinds of domains (smectic A and ferroelectric smectic $S_F$) coexist. The panel (d) seems to be the most spectacular one. It presents a well ordered nematic splay-bend structure, where in the absence of  positional order  the ribbon-like, splay-bend  modulation of  orientational order emerges. In the panel (e)  the splay-bend domains are observed for molecules of non-zero thickness. 

It should be noted that the  structures identified  in simulations,  along with  their thermodynamic properties, agree well with  predictions of density functional analysis. 
Even the periodicity ($\sim 10$ for needle boomerangs and $\sim 12$ for the HB particles -- see Fig.~\ref{Fig20_fig:theta}) and localization of the most disordered, splay-bend phase agree  with predictions of  bifurcation analysis  (see yellow region of  the bifurcation diagrams, Figs~(9, 12)).   
 It  proves that orientational order limited to two discrete orientations of the steric dipole with respect to the (local) director, which we  used for density functional analysis, allows for a 
 proper identification of the structures that can condense from the nematic phase 
 in the case of two-dimensional hard boomerangs.  However, there are also structures, which are not included in the presented bifurcation analysis, like the one in Fig.~(\ref{fig:MCsnapshots}) (f), where the molecules tend to self-organize by forming  oriented rectangles, or squared blocks, without any distinguishable layered structure. 

\section{Summary and conclusions}

We have studied two-dimensional ensembles  of  bent-core shaped molecules  of zero and finite arm width,
confined to the planar surface.
 Using the second virial Onsager density functional theory and the bifurcation analysis
 the role of excluded volume interactions in stabilizing
different structures and its influence on  local polarization
have been examined.

Onsager's theory reconstructs main conclusions from \cite{Gonzales} about
occurrence of the $S_{AF}$ phase and proves that $S_{AF}$
is indeed robust for 2D bent-core system. 
It also stays
in line with experimental observations of  Gong and Wan for banana-shaped
P-n-PIMB molecules absorbed onto a HOPG surface \cite{Gong}. This is in spite
of the fact that we disregarded any orientationally dependent interaction
between substrate and molecules ($V_{ext}=0$ in Eq. (5)), limiting the role
of the surface to confine molecules in 2D (assumption of strong planar anchoring).
That the surface can be considered smooth at the lengthscale of the molecular size
is justified by comparing the size of bent-core P-n-PIMB molecules (a few nanometers)
and the lattice spacing of HOPG  (~0.25nm). Also we should add that our dimensionless
shape parameter $\delta$ corresponds to $\delta \lesssim 0.2$  for P-n-PIMB.

The most interesting observation is the identification of the antiferroelectric  nematic $N_{SB}$  phase, which is stable 
for long bent-core molecules. This structure is foreseen from Onsager's theory and supported by exemplary, constant pressure
 MC simulation. To the best of our knowledge, it has not been reported  experimentally so far.

We find neither smectic C nor incommensurate smectic order
to become likely for these systems.
We show that the actual state  of the lamellar structures depend
strictly on the behaviour of the Fourier transforms of the appropriately
recognized  parts of the excluded volume.
According to this behaviour different transitions are plausible,
yet other phases than antiferroelectric smectic A
can be realized  for large packing fraction $\eta$.
In this limit, also  structures that are beyond the scope
of the Onsager approach, like glassy or crystalline ones,
can potentially form.

We show  that small structural modifications like the change of  the arm edges,
the apex angle, or thickness of the arm may
substantially influence the behaviour of the
order parameters, wave vector and even  phase diagrams.
We also demonstrate that the width
of the molecular arm influences the layer thickness.

The ordinary smectic A and smectic F phases
are expected to appear at high packing fractions $\eta$, Figs.~(18, 19).
Since $\eta>1$ in these cases, this rises an issue as whether such phase
should not be excluded on the ground that
Onsager’s DFT is formally justified  in the dilute gas limit (DGL).
The reason we believe  this is not the case is that
the mathematically similar form of the free energy as that of
Onsager's, Eqs.~(12, 13), can be obtained 
 by applying the Parsons-Lee (PL) rescaling/resummation technique \cite{Parsons,Lee}.
They showed that the effect of (infinite) hierarchy of higher-order virial terms can be partly taken
into account in (\ref{freeener}) by an appropriate renormalization of the second virial coefficient.
Operationally, the PL rescaling replaces the second-order virial packing
fraction, $\eta$ , entering Eq.~(\ref{freeener}) through $\bar{\rho}=\eta/S_{mol}$,
by an effective packing fraction, $\eta_{eff}$, which is a monotonic function of  $\eta$.
The PL procedure, developed essentially for 3D systems, has been  extended to 2D
by Varga and Szalai \cite{2000-Varga-Parsons-2dim}. One possibility, shown to work
well,  is equivalent to the replacement
\begin{equation}\label{parsons-convex}
  \eta_{eff} \rightarrow \frac{1}{2} \left(\frac{\eta}{1-\eta}-\log
   \left(1-\eta \right)\right).
\end{equation}
That is,  the physical range of $\eta \le 1$
is mapped  on the infinite region of  $\eta_{eff}\ge 0$.
Assuming, for example,  $\eta\lesssim 0.8 $
would  be equivalent  to substitute $\eta_{eff} \lesssim 2.8 $ in (\ref{freeener}).
Such rescaling of the free energy quantitatively improves the predictions
of Onsager's theory and shifts $S_A$ and $S_F$ to lower packing fractions.

Finally, we should mention that we carried out our calculations by assuming
that the reference nematic system is perfectly aligned.
Apparently this approximation seems to work quite well as indicated by exemplary 
MC simulation, which recover even the less ordered 
nematic splay-bend structure.

\section{Appendix A}
We add here, for completeness, the formulas for the Fourier transforms of the interaction-excluded slice kernel
for the SB boomerangs discussed in Section V.  Subscripts refer to coefficients calculated for cases shown in Fig.~(\ref{Fig10_fig:cases}).
%
\begin{equation}
\begin{array}{ll} \alpha_{I}(\psi, \delta, k) = \frac{1}{8k^2} & \left[ 4 \sec ^2(\psi )
\cos (2 k (\delta  \cot (\psi )-1)) \cos ^2(k-\delta  k \cot (\psi )) \right.  \\
&  \left. + \, \cos (2 \psi ) \sec ^2(\psi ) \cos (2 \delta  k \cot (\psi ))-2\cos (k (\delta  \csc (\psi ) \sec (\psi )-2)) \right.  \\
 & \left. +\,\, 2 \cos (2 k) \cos (2 \psi ) \sec ^2(\psi )-\cos (4 k) \left(\sec ^2(\psi )+2\right)-2 \right] \\
\end{array}
\end{equation}
\begin{equation}
\begin{array}{ll} \alpha_{II}(\psi, \delta, k) = \frac{1}{8k^2} & \left\{ - \tan (\psi )
\csc (2 \psi ) \left[2 \cos \left(4 k \cos ^2(\psi ) (\delta \cot (\psi )-1)\right)  \right. \right. \\
& \left. -\,\, 4 \cos (2 k (\delta  \cot(\psi )-1))-2 \cos (4 k (\delta  \cot (\psi )-1))+\cos (4 k-2 \psi ) \right. \\
& \left. \left. -\,\, 2 \cos (2 (k-\psi ))-2 \cos (2 (k+\psi ))+\cos (2 (2 k+\psi)) \right. \right. \\
& \left. \left. +\,\, 4 \cos (4 k)+2 \cos (2 \psi ) \right] \right\}
\end{array}
\end{equation}
\begin{equation}
\begin{array}{ll} \alpha_{III}(\psi, \delta, k) = \frac{1}{8k^2} & \left[ \sec ^2(\psi ) \left(2 \cos ^2(\psi ) \cos (2 k (\delta  \cot (\psi )-1))-2 \cos (2 \psi ) \cos (2 k (\delta  \cot (\psi )-2)) \right. \right. \\
& \left. \left. +\cos (4 k (\delta  \cot (\psi )-1))+\cos (4 k) (\cos (2 \psi )-2)\right) \right]
\end{array}
\end{equation}

\begin{equation}
\begin{array}{ll}
\beta_{I}(\psi, \delta, k) = \frac{1}{8 k^2} & \left[ \left(\sec ^2(\psi )-2\right) \cos (2
\delta  k \cot (\psi ))+2 \left(\cos (k (2-\delta  \csc (\psi ) \sec (\psi ))) \right. \right. \\
& \left. \left. +\, \sec^2(\psi ) \left(2 \cos (2 k (\delta  \cot (\psi )-1)) \sin ^2(k-\delta  k
\cot (\psi ))  \right. \right. \right. \\
& \left. \left.     +     \cos (2 k) \cos (2 \psi)\right)-1\right)
 \left. + \cos (4 k) \left(\sec ^2(\psi )-2\right) \right] \\
\end{array}
\end{equation}
\begin{equation}
\begin{array}{ll}
\beta_{II}(\psi, \delta, k) = -\frac{1}{16 k^2} & \left[ \sec ^2(\psi ) \left(-2 \cos \left(4 k
\cos ^2(\psi ) (\delta  \cot (\psi )-1)\right)-4 \cos (2 k (\delta  \cot (\psi )-1)) \right. \right. \\
& \left. \left. +\, 2 \cos (4 k (\delta  \cot (\psi )-1))+\cos (4 k-2 \psi )-2 \cos (2 (k-\psi )) \right. \right. \\
& \left. \left. - \,2 \cos (2 (k+\psi ))+\cos (2 (2 k+\psi ))+2 \cos (2 \psi )+4\right) \right]
\end{array}
\end{equation}
\begin{equation}
\begin{array}{ll}
\beta_{III}(\psi, \delta, k) = 
\frac{1}{8k^2} & \left[ \sec ^2(\psi ) \left(6 \cos ^2(\psi ) \cos (2 k (\delta  \cot (\psi )-1)) \right. \right. \\
& \left. \left. +\cos (2 \psi ) \left(\cos (4 k)-4 \cos ^2(k (\delta  \cot (\psi )-2))\right) \right. \right. \\
& \left. \left. -\cos (4 k (\delta  \cot (\psi )-1))-2\right) \right]
\end{array}
\end{equation}

\begin{equation}
\begin{array}{ll}
\beta_{I}(\psi, \delta, k \to 0) = \frac{1}{4} \csc (2 \psi ) \left(6 \delta ^2
\cos (2 \psi )+\left(\delta ^2-1\right) \cos (4 \psi )+3 \delta ^2-4 \delta  \sin (2 \psi
   )+1\right)
\end{array}
\end{equation}
\begin{equation}
\begin{array}{ll}
\beta_{II}(\psi, \delta, k \to 0) = \frac{1}{2} \sec ^2(\psi ) \left((\delta
\cot (\psi )-1)^2-2 \cos ^2(\psi ) \cot ^2(\psi ) (\sin (\psi )-\delta  \cos (\psi ))^2+\cos(2 \psi )\right)
\end{array}
\end{equation}
\begin{equation}
\begin{array}{ll}
\beta_{III}(\psi, \delta,k \to 0) = \frac{1}{2} \left(\delta ^2-2 \delta  \cot (\psi )+1\right)
\end{array}
\end{equation}
%
%
\section*{Acknowledgments}
This work was supported by Grant No. DEC-2013/11/B/ST3/04247 of the National Science Centre in Poland.

\end{document}